\documentclass[prl,preprint,url,superscriptaddress]{revtex4}

\usepackage[toc,page]{appendix}
\usepackage{graphicx,pdflscape,placeins}
\usepackage{graphicx,pdflscape}
\usepackage{color}
\usepackage{bm}
\usepackage{alltt,dsfont}
\usepackage{appendix}
\usepackage{amsmath,amssymb}
\usepackage{chngpage}
\usepackage{booktabs}
\usepackage{array}
\usepackage{multirow}
\usepackage{float}
\usepackage{chngpage}
\usepackage{natbib, graphicx}
\usepackage[caption=false]{subfig}
\usepackage{pifont}

\newcommand {\apgt} {\ {\raise-.5ex\hbox{$\buildrel>\over\sim$}}\ }
\newcommand {\aplt} {\ {\raise-.5ex\hbox{$\buildrel<\over\sim$}}\ }
\newcommand{\lessim}{\aplt}
\newcommand{\gssim}{\apgt}

\newcommand{\beq}{\begin{eqnarray}}
\newcommand{\eeq}{\end{eqnarray}}
\newcommand{\barray}{\begin{eqnarray}}
\newcommand{\earray}{\end{eqnarray}}
\newcommand{\nn}{\nonumber}

\newcommand{\disp}[1]{equation~(\ref{#1})} 

\newcommand{\refdisp}[1]{Ref.~(\onlinecite{#1})}
\newcommand{\figdisp}[1]{Fig.~(\ref{#1})}

\newcommand{\kk}{  \hat{k}}

\newcommand{\lv}{\text{V}_\text{L}}
\newcommand{\hv}{\text{V}_\text{H}}

\usepackage{hyperref}
\hypersetup{
    colorlinks,
    citecolor=red,
    filecolor=cyan,
    linkcolor=blue,
    urlcolor=magenta
}

\usepackage{atveryend}
\makeatletter
\let\origcitation\citation
\AtEndDocument{\def\mycites{}%
  \def\citation#1{\g@addto@macro\mycites{#1^^J}\origcitation{#1}}}
\AtVeryEndDocument{\newwrite\citeout\immediate\openout\citeout=\jobname.cit
 \immediate\write\citeout{\mycites}\immediate\closeout\citeout}
\makeatother

\begin{document}
\title{ Origin of  Kinks in  Energy Dispersion  of Strongly  Correlated Matter}
\author{ Kazue Matsuyama$^1$, Edward Perepelitsky$^{2,3}$ and B Sriram Shastry  }
\affiliation{ Physics Department, University of California, Santa Cruz, CA 95064 \\
$^2${Centre de Physique Th\'eorique, \'Ecole Polytechnique, CNRS, Universit\'e Paris-Saclay, 91128 Palaiseau, France}\\
$^3${Coll\`ege de France, 11 place Marcelin Berthelot, 75005 Paris, France}
 }
\date{\today}
\begin{abstract} %
We investigate  the origin  of ubiquitous  low energy kinks found in Angle Resolved Photoemission (ARPES) experiments in a variety  of correlated matter. Such kinks are unexpected from weakly interacting electrons and hence identifying  their origin should lead to fundamental insights in strongly correlated matter. We devise a protocol for extracting the kink momentum and  energy from the experimental  data which  relies solely on the  two asymptotic tangents of  each dispersion curve, away from the feature itself. It  is thereby insensitive to  the different  shapes of the kinks as seen in experiments. The body of available data is then   analyzed using this method. We proceed to discuss  two alternate theoretical explanations  of the origin of the kinks. Some  theoretical proposals  invoke local Bosonic excitations (Einstein phonons or other modes  with spin or charge character), located exactly at the energy of observed kinks, leading to a  {\em momentum independent} self energy of the electrons. A recent  alternate is the theory of  extremely correlated  Fermi liquids (ECFL). This theory  predicts kinks in the dispersion arising  from  a  {\em  momentum dependent}  self energy of correlated electrons. We present the essential results from both classes of theories, and  identify experimental features that can help distinguish between the two mechanisms. The ECFL theory is found to be consistent with currently available data on kinks in the nodal direction of cuprate superconductors, but conclusive tests require higher resolution energy distribution curve data.
 \end{abstract}
\maketitle

\section{Introduction} 
High precision measurements of electronic spectral dispersions has been possible in recent years, thanks to the  impressive  enhancement of the experimental resolution in the angle resolved photoemission spectroscopy (ARPES). This technique measures  the single electron spectral function $A(\vec{k},\omega)$  multiplied by  the Fermi occupation function; it can be scanned  at either fixed $\vec{k}$ as a function of $\omega$ or at fixed $\omega$ as a function of $\vec{k}$.  These scans produce respectively  the energy distribution curves  (EDCs) and momentum distribution curves (MDCs). The line shapes  in both these scans are of fundamental interest, since they provide a direct picture  of the quasiparticle and background components of interacting Fermi systems, and thus unravel the roles of various interactions that  are at play in strongly correlated  Fermi systems. The  dispersion relation of the electrons  can be studied  through the location of the  peaks of $A(\vec{k},\omega)$ in constant $\omega$ or constant $\vec{k}$ scans.

 Recent experimental  studies have displayed a surprising ubiquity of {\em kinks} in the dispersion of strongly correlated matter at low energies $\sim 50-100$ meV. The kinks are  bending type   anomalies (see \figdisp{kinkanatomy}) of the simple $\omega = v_F (\vec{k}-\vec{k}_F)$, i.e.   linear energy versus momentum dispersion that is expected near $\vec{k}_F$ from band theory.  The special significance of kinks lies in the fact that their existence {\em must} signal 
 a {\em departure} from band theory. This departure  could be either due to  electron-electron  interactions,  or to interaction of the electrons  with other Bosonic degrees of freedom. Either of them are therefore  significant enough to leave a direct and observable fingerprint in the spectrum.  The goal of this work is to elucidate the origin of the observed kinks, and   therefore  to throw light on the dominant interactions that might presumably    lead    to  high Tc superconductivity.

The purpose of this paper is multifold, we  (i)  survey the occurrence of the kinks in a variety of  correlated systems of current interest, (ii)  provide a robust protocol for characterizing  the kinks which is  insensitive to the detailed  shape of the kink, (iii)  discuss how these kinks arise  in two classes of theories, one based on coupling to a Bosonic mode and the other to strong correlations, and (iv)  identify testable predictions that  ARPES experiments can use to distinguish between these.

The fifteen systems reporting kinks are listed in Table~(\ref{survey}); these include (1) most high $T_c$ cuprates in the (nodal) direction $\langle11\rangle$ at various levels of doping from insulating to normal metallic states in the phase diagram  \cite{Zhou2003,Johnson2001}  (2) charge density wave systems, (3) cobaltates and (4) ferromagnetic iron surfaces. The kinks lose their sharpness as temperature is raised \cite{Lanzara2001,Johnson2001,Kaminski2001}, and appear to evolve smoothly between the d-wave superconducting state and the normal state. 
\begin{center}
 \begin{table}[htb]
 \centering
    \resizebox{\textwidth}{!}{\begin{tabular}{l l l l l l l c}
      \hline
        \multicolumn{1}{l}{Name of the compounds} &
        \multicolumn{4}{l}{} &
        \multicolumn{2}{l}{Local Bosonic Mode}&
        \multicolumn{1}{l}{}\\[-1em]
        \cline{1-1}\cline{6-8}
         &
        \multicolumn{2}{l}{Above Tc} &
        \multicolumn{2}{l}{Below Tc} &
        \multicolumn{2}{l}{}&
        \multicolumn{1}{l}{}\\[-1em]
         \cline{2-3}\cline{4-5}
        &
        \multicolumn{1}{l}{MDC} &
        \multicolumn{1}{l}{EDC} &
        \multicolumn{1}{l}{MDC} &
        \multicolumn{1}{l}{EDC} &
        \multicolumn{1}{l}{Charge}&
        \multicolumn{1}{l}{Spin}&
        \multicolumn{1}{l}{Not reported}\\[-1em] 
         \cline{2-2}\cline{3-3}\cline{4-4}\cline{5-5}\cline{6-6}\cline{7-7}\cline{8-8}       
        LSCO   & \checkmark\cite{Lanzara2001, Garcia2010} &  & \checkmark\cite{Lanzara2001, Zhou2003, Yoshida2007, Garcia2010} & \checkmark\cite{Mishchenko2011} & \checkmark\cite{McQueeney1999, Pintschovius1999, Fukuda2005}&\checkmark\cite{Vignolle2007} &\\[-1em]
        Bi2201 & \checkmark\cite{Lanzara2001, Garcia2010, Sato2003, Yang2006, Ying2013} & \checkmark\cite{Meevasana2007}  &  \checkmark\cite{Sato2003, Ying2013} & &\checkmark\cite{Graf2008}& & \\[-1em]
        Bi2212   & \checkmark\cite{Bogdanov2000, Johnson2001, Lanzara2001, Kaminski2001,Sato2003, Garcia2010, Zhang2008}   & \checkmark\cite{Kaminski2001}  & \checkmark\cite{Bogdanov2000, Johnson2001, Lanzara2001, Kaminski2001, Sato2003, Garcia2010, Zhang2008}   &   &\checkmark\cite{Vig2015} & \checkmark \cite{Fong1999, He2001}& \\[-1em]
        Bi2223 & \checkmark\cite{Sato2003, Ideta2008}   &   & \checkmark\cite{Sato2003, Ideta2008, Ideta2013}  &   &  & &\checkmark \\[-1em]
        YBCO   &  &   & \checkmark\cite{Borisenko2006}  &   &  \checkmark\cite{Reuchardt1989, Pintschovious2004}& \checkmark\cite{Rossat1991, Mook1993, Dai1996, Dai2001}& \\[-1em]
        Hg1201   &  &   & \checkmark\cite{Vishik2014}  &   & \checkmark\cite{dAstuto2003} &\checkmark\cite{Li2010, Li2012, Chan2016} &  \\[-1em]
        F0234   &  &   & \checkmark\cite{Chen2009}  &   &    & & \checkmark \\[-1em]
        CCOC   &  &  & \checkmark\cite{Ronning2003}  &  &    & &\checkmark \\[-1em]
        LSMO   &  &   & \checkmark\cite{Mannella2005}  & \checkmark\cite{Mannella2005} &   & &\checkmark \\[-1em]
        2H-TaSe2 (CDW) &  &   & \checkmark\cite{Valla2000}  &  &\checkmark\cite{Brusdeylins1990}  & & \\[-1em]
        Iron (110) surface  &  &   & \checkmark\cite{Schafer2004} 85 K  &  &    & &\checkmark \\[-1em]
        BiBaCo1   &  &   & \checkmark\cite{Brouet2012} 5K  & \checkmark\cite{Brouet2012} 5K  &   & &\checkmark \\ [-1em]
        BiBaCo2  &  &   & \checkmark\cite{Brouet2012} 5K  & \checkmark\cite{Brouet2012} 5K &    & &\checkmark \\[-1em]
        BiBaCo   & \checkmark\cite{Brouet2012} 200K & \checkmark\cite{Brouet2012} 200K  &   &  &    & &\checkmark \\[-1em]
        NaCoO  &  &   & \checkmark\cite{Brouet2012} 5K  & \checkmark\cite{Brouet2012} 5K  &   & &\checkmark \\ \hline
    \end{tabular}}    
    \caption{Comprehensive survey for ARPES kinks}
    \label{survey}
   \end{table}
  \end{center}
\FloatBarrier
The kinks above $T_c$ are smoothed out as one  moves away from nodal direction \cite{Sato2003}.  Recent experiments \cite{He2013} resolve this movement of the kinks more finely into two sub features. Most  of the  studies in Table~(\ref{survey}) focus on MDC kinks,  the EDC kinks data is available for only eight systems so far.  Bosonic modes have been reported in six systems using different probes such as inelastic x-rays or magnetic scattering,  with either charge (phonons, plasmons)  or spin (magnetic)  character,  while the remaining nine systems  do not report such modes.  
A few theoretical studies of the kinks  have implicated the observed low energy modes via  electron-Boson type  calculations;  we summarize   this  calculation  in the Supplementary Information (SI) \cite{supplement}. We find,  in agreement with earlier studies, that  the Boson coupling  mechanism  yields kinks in the MDC dispersion, provided  the electron-Boson coupling is taken to be sufficiently large. In addition, we find   in all cases studied, this mechanism    also  predicts   a jump in the EDC dispersion. It also   predicts an extra   peak in the spectral function pinned to the kink energy after the wave vector crosses the kink. These two features are experimentally testable and differ from  the predictions of the correlations mechanism discussed next.

Since kinks are also observed in cases  where no obvious Bosonic mode is visible, it is important to explore  alternate mechanisms that give rise to such features.  In this context we note that a recent theoretical  work using    the  extremely strongly correlated Fermi liquid (ECFL)   theory \cite{Shastry-2011, Shastry2014} calculates the dispersion using a low momentum and  frequency expansions of the constituent self energies. This calculation \cite{Shastry2014} shows that both  EDC and MDC energy dispersions  display qualitatively similar kinks, in particular there is no jump in either  dispersion. In essence this work implies  that a purely electronic mechanism with a strong momentum dependence of the Dyson self energy results in  kink type anomalies. In terms of parameter counting, the calculation is {\em  overdetermined}, it can be represented  in terms of four  parameters which   can be fixed from a subset of measurements. With this determination one can  then predict many other measurables and testable relations between these- as we show below.    We show below that the various  predictions are  reasonably satisfied in one case (of OPT Bi2212 below), while   in  other cases, there is insufficient   experimental data to test the theories.
 
The ECFL theory incorporates strong Gutzwiller type correlation effects into the electron dynamics \cite{supplement}. It produces line shapes that  are  in close correspondence to experimental results for the high $T_c$ systems \cite{Gweon2011,Matsuyama2013}. The presence of a low energy kink in the theoretical  dispersion was already noted in \refdisp{Gweon2011}, the present work substantially elaborates this observation. In order to understand the origin of a low energy scale  in the ECFL theory, it is useful to recall the predicted  cubic correction to Fermi liquid self energy  $\Im m \, \Sigma(\vec{k}_F,\omega) \sim \omega^2(1- \frac{\omega}{\Delta_0})$ from equations~(SI-42, \ref{caparison},\ref{spectral-function}).  Here $\Delta_0$ is an  emergent low energy scale, it is related to the   correlation induced reduction of the quasiparticle weight $Z$.  It  reveals  itself   most clearly in the observed particle hole asymmetry of the spectral functions, and therefore  can be estimated independently from spectral {\em  lineshape} analysis. A related and  similar low value of the effective Fermi temperature is found in recent studies of the resistivity \cite{Shastry-2016}.   Here and   in our earlier studies it is coincidentally found that  $\Delta_0\sim20-50$ meV, i.e. it is also  roughly  the  energy scale of the kinks when the bandwidth is a few eV.
 
\section{ ARPES spectral dispersions, kinks and a protocol for   data analysis} 
\subsection{Summary of  variables in the theory}
\begin{figure}[htb]
\begin{center}
\includegraphics[width=0.6\columnwidth]{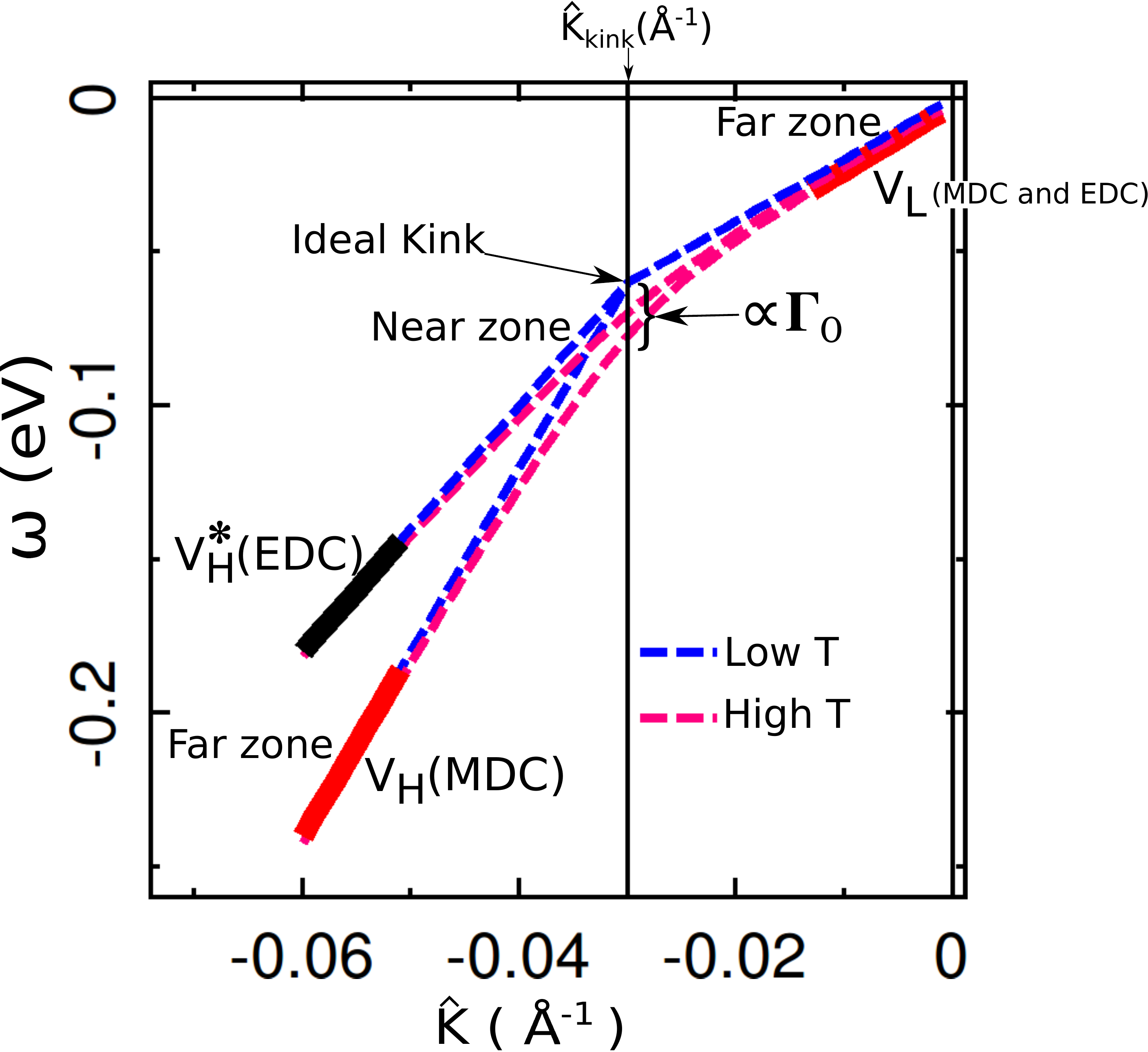}
\end{center}
\caption{  A schematic  MDC and EDC spectrum   displaying typical features  of experiments discussed below.  Here   $\kk=  (\vec{k}- \vec{k}_F). \vec{\nabla}\varepsilon_{k_F}/|\vec{\nabla}\varepsilon_{k_F}|$, is the momentum component normal to the Fermi surface, and we label EDC variables with a star. (The sketch uses  parameters $\lv$ = 2\,eV\,$\AA$, $\hv$ = 6\,eV\,$\AA$, $r = 1.5, \hat{k}_{kink} = - 0.03\,\AA^{-1}$, $\Delta_0 = 0.03$\,eV, and $\Gamma_0$ = 0.01\,eV in equation~(\ref{EDC},\ref{MDC})).
The tangents in the {\em far zones}  identify the  asymptotic velocities $V_L< V_H$ and $V^*_L< V^*_H$ 
that characterize the MDC and EDC spectra. The intersection of the extrapolated  MDC tangents fixes the  kink momentum $\kk_{kink}$  and the ideal energy $E^{ideal}_{kink}$. The dispersion is rounded with raising T, as in the lower (red) curve. We {\em define} the MDC kink energy $E^{MDC}_{kink}$ as $E(\hat{k}_{kink})$, i.e.  the binding energy measured {\em at} the kink momentum, and similarly the EDC kink energy.  In all cases     $V_L=V^*_L$.  
A testable consequence of  the   ECFL   theory is that   $V_H^*$ is fixed  in terms of the two MDC velocities by a strikingly simple relation:   $V^*_H= \frac{3 V_H -V_L}{V_H+V_L} \times V_L $. This prediction is tested  against experimental data in  \figdisp{fig2} where both EDC and MDC data is available. In contrast the electron-Boson theory predicts a jump in the EDC dispersion at the kink energy, followed by $V_H^*=V_H$. Note that the difference between the EDC (MDC) kink energy, $E^{EDC}_{kink} = E^{ideal}_{kink} - \Gamma_0$ and $E^{MDC}_{kink} = E^{ideal}_{kink}- \Gamma_0 \sqrt{\frac{r}{2-r}}$, and the ideal kink energy is equal (proportional) to $\Gamma_0$}
\label{kinkanatomy}
\end{figure}

 A few common features of spectral dispersions   found in experiments   are summarized in  \figdisp{kinkanatomy}.   The schematic figure  shows a region of low spectral velocity near the Fermi level followed by a region of steeper velocity, these are separated by a bend in the dispersion-  namely the kink. While the kink itself has a somewhat variable shape in different experiments, the ``far zone'' is much better  defined and is usually independent of the temperature, we denote the velocities in the far zones $V_L,V_H$  for the MDC dispersion and the EDC dispersion counterparts by $V_L^*,V_H^*$. In terms of the normal component of the  momentum measured from the Fermi surface  
 \beq \hat{k}= (\vec{k}- \vec{k}_F). \vec{\nabla}\varepsilon_{k_F}/|\vec{\nabla}\varepsilon_{k_F}|, \label{fermi-mon} \eeq  the kink momentum $\hat{k}_{kink}$  is uniquely defined  by extrapolating the two asymptotic tangents, and the binding energy at this momentum defines the ideal kink energy $E^{ideal}_{kink}$ (see \disp{ideal}), which serves as a useful  reference energy.

Our picture is that all lines of temperature varying MDC dispersion curves in near zone converges into one line in the far zone in \figdisp{kinkanatomy}. We find that both the  low and high velocities  are independent of  the temperature while depending on the doping levels. Lastly, the new laser ARPES data reveals that we need low temperature dispersion data to determine $V_L$ because temperature effect strongly influences the spectrum  near the Fermi level.

We first  define the important  ratio parameter $r$ ($1\leq r\leq 2$) from the MDC dispersion velocities as 
\beq
r= \frac{2 V_H}{V_H+V_L}. \label{r}
\eeq
  The   EDC  dispersion relation $E^*(\hat{k})$  locates the maximum of the spectral function $A(\vec{k},\omega)$ in $\omega$ at constant $\hat{k}$,   while   the MDC dispersion and $E(\hat{k})$ locates the maximum $\hat{k}$ at a fixed energy $\omega$. These    are found from the ECFL theory (see SI \cite{supplement} and \refdisp{Shastry2014}) as:
 \beq
 E^*(\hat{k})&=&  \left(  r \, V_L \hat{k} + \Delta_0 - \sqrt{\Gamma_0^2+ Q^2} \right),\label{EDC}  \\
 E(\hat{k})&=& \frac{1}{2-r} \left(  V_L \hat{k} + \Delta_0 - \sqrt{r (2-r) \, \Gamma_0^2  + Q^2} \right)  \label{MDC}, 
 \eeq
 where we introduced an energy parameter related to $r,V_L$ and $\hat{k}_{kink}$  
\beq
\Delta_0 = \hat{k}_{kink} V_L (1-r),  \label{delta0}
\eeq
 and a momentum type variable $Q= (r-1) \, V_L \, (\hat{k}- \hat{k}_{kink})$.
The  variable $\Gamma_0$ is temperature like,
\beq
 \Gamma_0 = \eta + \pi \{ \pi k_B T\}^2/\Omega_\Phi; \label{rdef}
\eeq
here $\eta$ is an elastic scattering parameter  dependent upon the incident  photon energy,  it is very small for laser ARPES experiments and can be neglected to a first approximation. Here  $\Omega_\Phi$ is a self energy decay constant  explained further in the SI \cite{supplement}.  
The ideal kink energy $V_L \hat{k}_{kink}$ can be expressed in terms of $\Delta_0$ scale as:
\beq
E^{ideal}_{kink}= - \frac{1}{r-1} \Delta_0 \label{ideal}.
\eeq

It is important to note that these dispersion relations equations~(\ref{EDC},\ref{MDC}) are different from the standard dispersion relations $E_{FLT}(\hat{k})=E_{FLT}^*(\hat{k})= V_H \hat{k}$, which follow  in the simplest  Fermi Liquid Theory (FLT) near the Fermi energy $A_{FLT}(\vec{k},\omega)= \frac{1}{\pi} \frac{\Gamma_0}{(\omega-V_H \hat{k})^2+\Gamma_0^2}$. The  FLT dispersions are identical in EDCs and MDCs, and are independent of the temperature-like variable $\Gamma_0$, and do not  show kinks. On the other hand equations~(\ref{EDC},\ref{MDC}) do have kinks- as we show below, and the temperature-like variable $\Gamma_0$ plays a significant role in the dispersion. At $\Gamma_0=0$ one has an ideal spectrum, where the kinks are sharpest. When $\Gamma_0\neq 0$, due to either finite temperature or finite damping $\eta$,  related to   the  energy of the incoming photon,  the kinks are rounded.

A few consequences of equations~(\ref{EDC},\ref{MDC}) can be noted for the purpose of an experimental determination of the Fermi momentum. The chemical potential is usually fixed by referencing an external metallic contact and is unambiguous. Experimentally the Fermi momentum is usually found from the MDC, as the momentum where the spectral function is maximum with energy  fixed at the  chemical potential, i.e. $\omega=0$. This corresponds to  the generally wrong  expectation, that   $E(\hat{k}_{peak})=0$ implies $\hat{k}_{peak}=0$.
When $\Gamma_0 \geq 0 $, from \disp{MDC} we see that the condition $E(\hat{k}_{peak})=0$  gives $\hat{k}_{peak}= \frac{\sqrt{\Delta_0^2+ r^2\,\Gamma_0^2}- \Delta_0}{ r V_L},$ a positive number that equals zero only in the ideal case $\Gamma_0=0$. Thus there is an apparent enlargement of the Fermi surface due to a finite $\Gamma_0$ that needs to be corrected. By the same token, at the true (Luttinger theorem related) Fermi momentum $\hat{k}=0$,  the MDC energy  $E(0)= \frac{\Delta_0 - \sqrt{\Delta_0^2+ r(2-r) \Gamma_0^2}  }{2-r}$, a  negative number when $\Gamma_0 \neq 0$.  In recent laser ARPES Bi2201 data  \refdisp{Ying2013} (panel (a) in Fig. (4)), we see that $E(\hat{k}_{peak})$ vanishes at increasing $\hat{k}_{peak}$ as T is raised, as predicted in our calculation.  Recent laser ARPES experiment on OPT Bi2212 compounds reports a  similar temperature dependence of momentum of MDC dispersion at the Fermi level in \refdisp{Zhang2008}, strongly supporting our picture of its origin.

Similarly, the EDC  peak at the true Luttinger theorem related  Fermi surface  $\hat{k}=0$ is non-zero. We find $E^*(0)=\left(  \Delta_0-\sqrt{\Delta_0^2+\Gamma_0^2}\right) \leq 0 $. Clearly   $E^*(0)$ is  negative unless  $\Gamma_0 = 0$, i.e. it is generically red-shifted. If we are tempted to  identify the Fermi momentum   from the  condition $E^*(\hat{k}^*_{peak})=0$, a similar cautionary remark is needed.
The condition  $E^*(\hat{k}^*_{peak})=0$ gives $\hat{k}^*_{peak}= \frac{\sqrt{\Delta_0^2+ (2 r-1)\Gamma_0^2}- \Delta_0}{ (2 r-1) V_L}, $ again a  positive number as in the MDC case, and thus a slightly different enlargement of the apparent Fermi surface.

The above comments illustrate the difficulty of finding the correct Fermi surface when $\Gamma_0$ is non-negligible, as in the case of synchrotron ARPES with substantial values $\Gamma_0\gssim 50$meV. On the other hand the laser ARPES studies have a much smaller $\eta\lessim 10$ meV, where our analysis can be tested by varying the temperature and the consequent change of the spectrum. In the following, we analyse the data from the Bi2201 system where the laser data is available at various T, and allows us to test the above in detail. Our analysis below of two other synchrotron data, the OPT Bi2212 has  $10 \leq \eta \leq 40$ meV,  while the low T LSCO data is assumed to be in the limit of $\eta = 0$ because of the lack of high temperature dispersion data.

\FloatBarrier
The spectral function at low frequencies close to $\vec{k}_F$ is also obtainable from these parameters, the relevant formula is noted below. In terms of $\xi$
\beq
 \xi&=& \frac{1}{\Delta_0} (\omega- r\,V_L \hat{k} ) \label{caparison}
\eeq
 the spectral function is:
\beq
A(\vec{k},\omega)= \frac{z_0}{\pi} \frac{\Gamma_0}{(\omega- V_L \hat{k} )^2 + \Gamma_0^2} \times  \{ 1-  \frac{\xi}{\sqrt{1+ c_a \xi^2} }  \}, \nn \\ \label{spectral-function}
\eeq
Here $z_0$ is the quasiparticle weight and $c_a\sim 5.4$  (see SI \cite{supplement}).  We should keep in mind  that  these expressions follow from   a  low energy expansion, and  is limited to small $\hat{k}$ and $\omega$; in practical terms  the dimensionless variable $|\xi| \lessim 4$, so that $\omega$ (or $\hat{k}$) is bounded by the kink energy (or momentum), as defined  below.

\section{OPT Bi2212 ARPES dispersion data}
\begin{figure}[htp]
\begin{center}
\includegraphics[width=0.9\columnwidth]{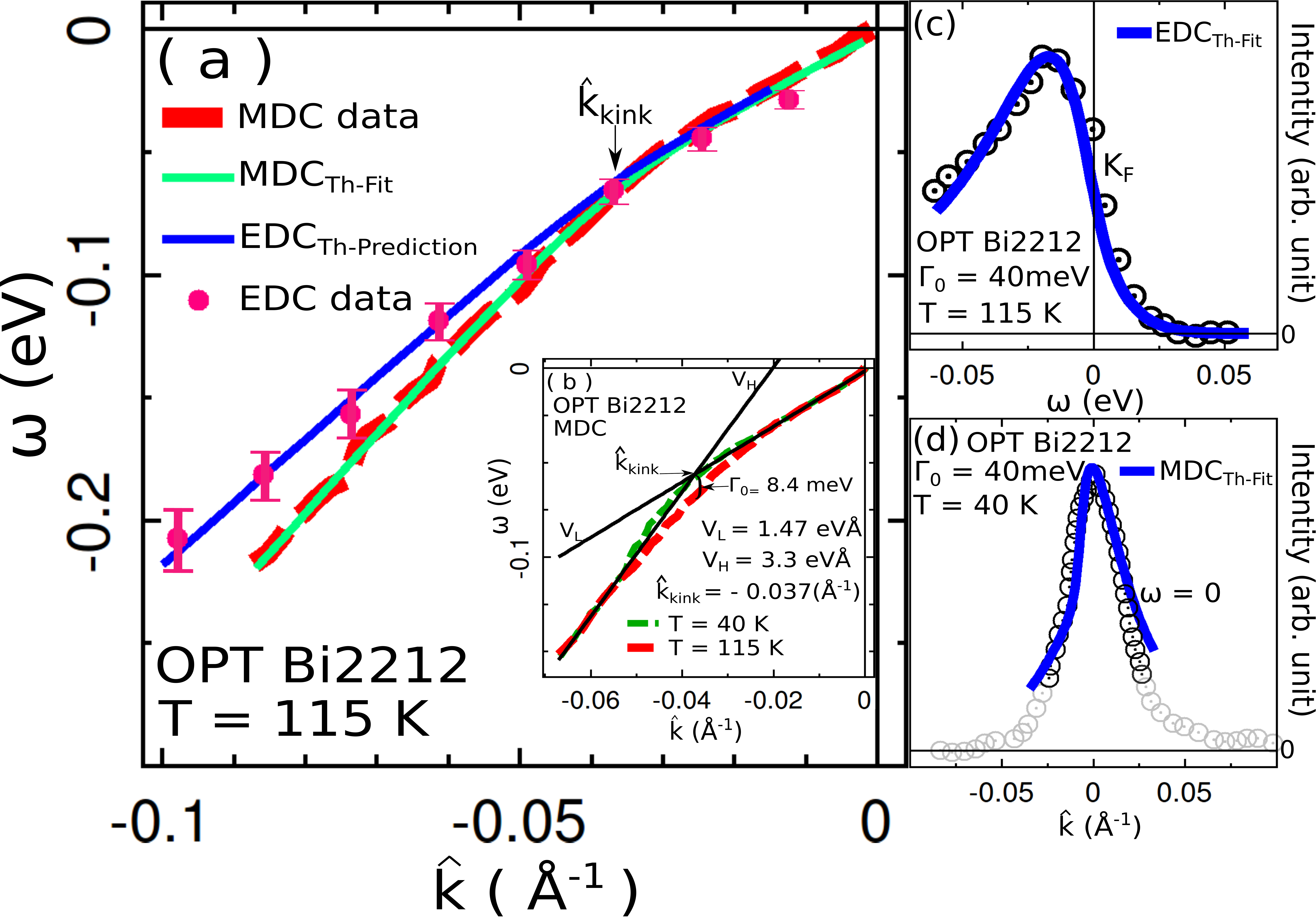}
\end{center}
\begin{minipage}{\textwidth}
\caption{\footnotesize ARPES kinks data for OPT Bi2212 from  \refdisp{Kaminski2001} compared to theoretical ECFL curves (solid lines) using  parameters listed in  Table.~\ref{table1}.
 {\bf Panel:(a)}  The { predicted}   EDC spectrum (blue)  from \disp{EDC}, versus the experimental EDC data (magenta symbols) at T=115K. For reference we also show the MDC data (red dashed curve) and the corresponding ECFL fit (green solid curve). {\bf Panel:(b)} Experimental MDC spectra at 40K (below $T_c$ in green dashed line) and 115K (above $T_c$ in red dashed line) yield common asymptotes shown in black lines  from the far zone. These determine the parameters displayed in Table~(\ref{table1}). {\bf Panel:(c)} At low energy $\pm$ 60 meV, the   EDCs spectral function (blue solid line) from \disp{spectral-function}  is contrasted with the corresponding ARPES data from \cite{Kaminski2001}.  {\bf Panel:(d)} At $\omega=0$  we compare the MDCs spectral function (blue solid line) from \disp{spectral-function}  with the corresponding ARPES data from \refdisp{Kaminski2001}. The range of validity for the theoretical expansion  is $\pm\,\hat{k}_{kink}$\,(\,0.037$\AA^{-1}$\,),  the  data points in the range  are shown in black circle symbols, while the  light gray circle symbols are  outside  this range. The peak position of the theoretical curve has been shifted to left by 0.007\,$\AA^{-1}$, a bit less than the instrumental resolution. A similar shift is made in Panel (l)  Fig.~\ref{LSCOEDC}. For analogous reasons the EDC peak in $A(k,\omega)$  at $\vec{k}_F$   is shifted to the left i.e. $E^*(0) \leq 0$. A small shift to the {\em right} is made in Panel (k) of \figdisp{LSCOEDC}, in order to compensate for this effect. These shift effects are within the resolution with present setups, but should be interesting to look for  in future generation experiments, since they give useful insights into the energy momentum dependence of the spectral function.
}
\label{fig2}
\end{minipage}
\end{figure}

In the well studied case of optimally doped Bi2212 (BSCCO) superconductors,   the kink has been observed in both EDC and MDC.
We summarize the ECFL fit parameters in Table~(\ref{table1}) obtained  from literature \cite{Kaminski2001}. We also display the predicted energy and high velocity of the EDC dispersion. The velocity ratio $V_H/V^*_H\sim 1.3$ in this case, is quite large and measurable.   In this case the EDC dispersion has fortunately already been measured, allowing us to   test the prediction.   From Table~(\ref{table1}) we see that the energy of the EDC kink and its velocity are close to the predictions.\\
 \begin{widetext}
\begin{table}[H]
\begin{center}
\resizebox{\textwidth}{!}{\begin{tabular}{| c c c ||c c|c  c||c c|}
   \hline
   \multicolumn{5}{|c |}{MDCs}& \multicolumn{4}{c| }{EDCs}\\\hline
   \multicolumn{3}{| c }{OPT Bi2212 ARPES data} &\multicolumn{2}{c |}{$E^{MDC}_{kink}$(meV)}&\multicolumn{2}{c}{$E^{EDC}_{kink}$(meV)}&\multicolumn{2}{c |}{$V_H^*$ (eV\,$\AA$)}\\
   \hline
$V_L$ (eV\,$\AA$) & $V_H$ (eV\,$\AA$) & $\hat{k}_{kink}$ ($\AA^{-1}$) & {\footnotesize Calculated}& {\footnotesize Measured} &{\footnotesize Calculated} & {\footnotesize Measured} & {\footnotesize Predicted} & {\footnotesize Measured} \\ \hline 
1.47 \,$\pm$\,0.07  & 3.3\,$\pm$\,0.3 &  -\,0.037\,$\pm$\,0.005 & 67\,$\pm$\,21  & 67\,$\pm$\,8 & 63\,$\pm$\,21 & 65\,$\pm$\,8 & 2.60\,$\pm$\,0.56 & 2.1\,$\pm$\,1.1\\ 
 \hline
\end{tabular}}
\begin{minipage}{\textwidth}
\caption{\small Parameter table for ARPES kink analysis for OPT Bi2212 \cite{Kaminski2001} in Fig.~\ref{fig2} presents three essential parameters, $V_L$, $V_H$, and $\hat{k}_{kink}$. From the high and low temperature MDC dispersions, we measured $\Gamma_0 \lesssim 10$ meV in Panel (b) of Fig.~\ref{fig2}. With the measured experimental parameters and determining the velocity ratio r in \disp{r}, we are able to estimate the finite temperature kink energy for EDC and MDC dispersions by $E^{EDC}_{kink} =  E^{ideal}_{kink} - \Gamma_0$ and $E^{MDC}_{kink} = E^{ideal}_{kink}- \Gamma_0 \sqrt{\frac{r}{2-r}}$ and predict $V_H^*$ by $V^*_H= \frac{3 V_H -V_L}{V_H+V_L} \times V_L$. The uncertainties for calculated variables were determined by error propagation, and the uncertainties for experimental variables were given by the half of the instrumental resolution.}
\label{table1}
\end{minipage}
\end{center}
\end{table}
\end{widetext}
In  Panel (a) in Fig.~\ref{fig2}, we plot the predicted EDC dispersion using the parameters extracted from the MDC dispersion in Panel (b), and compare with the ARPES data measured\cite{Kaminski2001}. It is interesting that the predicted slope of the  EDC  dispersion from $V^*_H= \frac{3 V_H -V_L}{V_H+V_L} \times V_L$ is close to the measured one. Indeed the measured EDC dispersion is close to that expected from the ECFL theory.  To probe  further, in Panel (c) in \figdisp{fig2}  we   compare the theoretical EDC line shape (solid blue line) given by the same parameters through \disp{spectral-function}, with the  ARPES line shape measured at high temperature \cite{Kaminski2001}. Panel (d) compares the theoretical MDC curve with the data. The theoretical curves are from the low energy expansion and hence are chopped at the high end, corresponding to roughly $|\xi|_{max}\sim \frac{r\,\lv\,\hat{k}_{kink}}{\Delta_0}$ for MDC and $|\xi|_{max}\sim \frac{E^{ideal}_{kink}}{\Delta_0} $ for the EDC. With this cutoff, the momentum is less than the kink momentum and the energy is less than the kink energy. We used $\Gamma_0$ = 40 meV  since it provides a rough fit for both EDC and MDC spectral functions.  
\FloatBarrier
This value is somewhat larger than the bound $\sim 10$ meV given in Table~(\ref{table1}), a smaller value leads to narrower lines but with the same shape.  In rigorous terms the same $\Gamma_0$ must fit the dispersion and also the spectral functions. Our fit, requiring a different  $\Gamma_0$, is  not ideal in that sense. However  the resolution of the available data is somewhat rough, and should  improve with the newer experimental setups that have become available. We thus expect that higher resolution data with laser ARPES should provide an interesting challenge to this theory.  We also  stress  that from  \disp{spectral-function}, the MDC line shapes look more symmetric than the EDC line shapes at low energies. While many  experimental results do show rather symmetric MDCs, there are well known exceptions.  For instance MDCs asymmetry has indeed been reported for nearly optimally doped Hg1201 (\,Tc = 95 K\,) at binding energy very close to the Fermi level, $\omega\sim$\,- 5 meV and  $\omega\sim$\,-18 meV in Fig.~5 in \refdisp{Vishik2014}. Note that the $\omega=0$  MDC  plot  of the spectral function $A(k,\omega)$ from \disp{spectral-function},  locates the peak momentum $\hat{k}_{peak}> 0$, i.e.  slightly to the right of the physical Fermi momentum $\vec{k}_F$, and we consider this implies that the {\em experimental } Fermi momentum determination is subject to such a correction, whenever  the spectral function \disp{spectral-function} has a momentum dependent caparison factor (see caption in \figdisp{fig2}).

\section{LSCO low temperature data}
\begin{figure}[htb]
\includegraphics[width=\columnwidth]{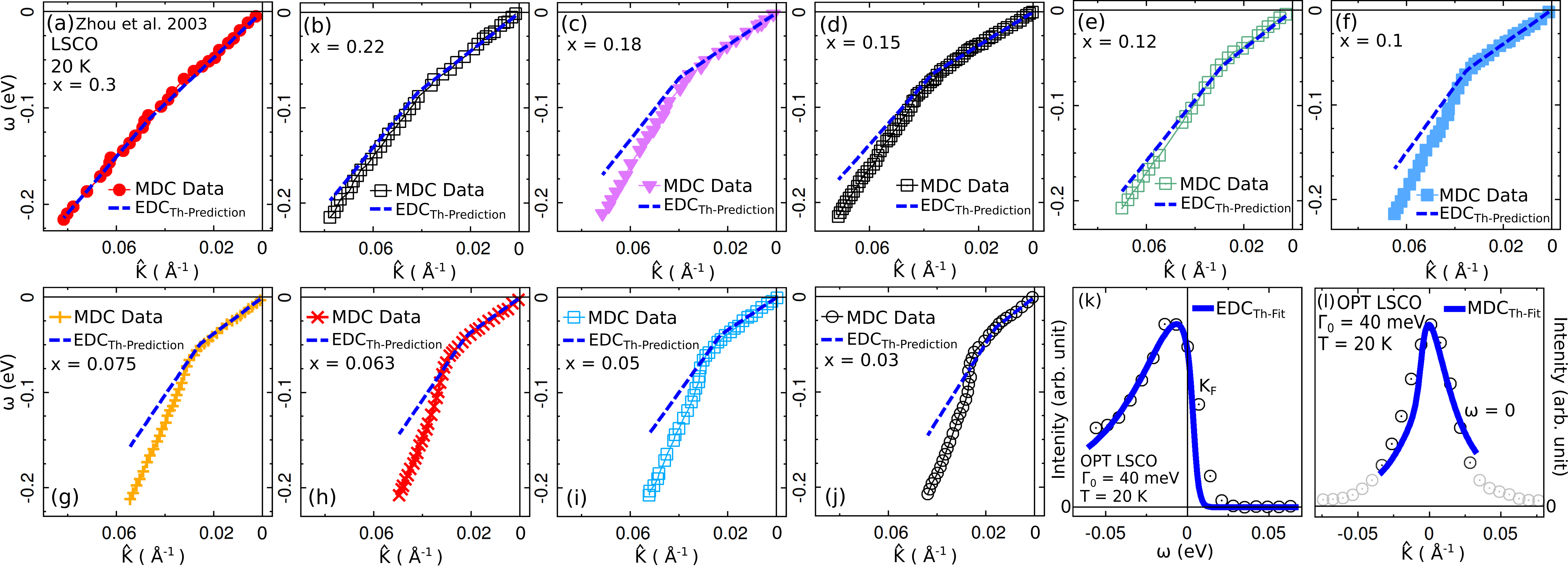}
\begin{minipage}{\textwidth}
\caption{ ARPES kinks data for LSCO data \cite{Zhou2003}  compared to 
theoretical ECFL curves (solid lines) using  parameters listed in  Table.~\ref{table2}.
The doping level x  varies between (normal metal) $0.3 \leq$ x $\leq 0.03$ (insulator)  in Panels   (a) to (j). Each panel shows MDC nodal dispersion  data (symbols), whose  uncertainties are  $\pm$ 10 meV. The blue dashed line is the theoretical prediction for EDC dispersion by \disp{EDC}. 
{\bf Panel:(k)} We compare the spectral line shape for EDCs at $k_F$  from \disp{spectral-function} (blue solid line) in the range  $\pm\,E^{ideal}_{kink}$ $\sim$ \,65 meV with  the corresponding ARPES data (black circles) \cite{Matsuyama2013}. 
 {\bf Panel:(l)}  At $\omega=0$  we compare the MDCs spectral function (blue solid line) from \disp{spectral-function}  with the corresponding ARPES data from \refdisp{Matsuyama2013}. The range of validity for the theoretical expansion  is $\pm\,\hat{k}_{kink}$\,(\,0.037$\AA^{-1}$\,),  the  data points in the range  are shown in black circle symbols, while the  light gray circle symbols are  outside  this range. The peak position of the theoretical curve MDC has been shifted to left by 0.006\,$\AA^{-1}$.} 
\label{LSCOEDC}
\end{minipage}
\end{figure}

Here we analyze the LSCO data at low temperature (20 K)  and at various doping levels raging  from the  insulator (x = 0.03) to normal metal (x = 0.3) from \refdisp{Zhou2003}. The parameters are listed in Table~(\ref{table2}), where we observe that the velocity $V_L$ is roughly independent of $x$, and  has a somewhat larger magnitude to that in OPT Bi2212 in Table~(\ref{table1}). The kink momentum decreases  with decreasing  x, roughly as $\hat{k}_{kink} = -(0.37 x-0.77 x^2) \AA^{-1}$, and the kink energies of EDC and MDC dispersions are essentially identical. In the region beyond the kink, the prediction for $V_H^*$  is interesting since it differs measurably  from the MDC  velocity $V_H$.  We find the ratio $V_H/V_H^* \sim 1.02-1.5$ is quite spread out  at different doping.
\begin{table}[htb]
\begin{center}
   \resizebox{\textwidth}{!}{\begin{tabular}{|c|c|c|c||c c|c c||c c|} 
   \hline
   \multicolumn{6}{|c}{MDCs} &\multicolumn{4}{| c |}{EDCs}\\
   \hline
   \multicolumn{4}{| c ||}{LSCO low temperature ARPES data} & \multicolumn{2}{c |}{$E^{MDC}_{kink}$(meV)}&\multicolumn{2}{c ||}{$E^{EDC}_{kink}$(meV)} &\multicolumn{2}{c |}{$V_H^*$ (eV\,$\AA$)}\\
   \hline
 x\,(\,{\footnotesize doping level}\,) & $V_L$ (eV\,$\AA$) & $V_H$ (eV\,$\AA$) & $\hat{k}_{kink}$ ($\AA^{-1}$) &Calculated & Measured & Calculated & Measured &Calculated & Measured\\ 
 \hline
   0.3& 2.4\,$\pm$\,0.2 &  3.0\,$\pm$\,0.3 & -\,0.047\,$\pm$\,0.005 & 113\,$\pm$\,29 & 110\,$\pm$\,10 &113\,$\pm$\,29 & &2.93\,$\pm$\,0.45 &\\ \hline
   0.22& 2.0\,$\pm$\,0.1 & 3.6\,$\pm$\,0.2 & -\,0.042\,$\pm$\,0.005 & 84\,$\pm$\,18 &85\,$\pm$\,10 & 84\,$\pm$\,18& &3.14\,$\pm$\,0.35&\\ \hline
   0.18& 1.7\,$\pm$\,0.3 & 4.5\,$\pm$\,0.6 & -\,0.040\,$\pm$\,0.005 & 68\,$\pm$\,43 &72\,$\pm$\,10  & 68\,$\pm$\,43 & &3.2\,$\pm$\,1.2 &\\ \hline
   0.15& 1.75\,$\pm$\,0.07 & 4.3\,$\pm$\,0.1  & -\,0.037$\pm$\,0.005 & 65\,$\pm$\,11 & 64\,$\pm$\,10 & 65\,$\pm$\,11& &3.23\,$\pm$\,0.20 & \\ \hline
   0.12& 2.0\,$\pm$\,0.3 & 3.7\,$\pm$\,0.5  & -\,0.029\,$\pm$\,0.005 & 58\,$\pm$\,28 & 55\,$\pm$\,10 & 58\,$\pm$\,28 & &3.19\,$\pm$\,0.89 &\\ \hline
   0.1& 1.8\,$\pm$\,0.2 & 5.0\,$\pm$\,0.7 & -\,0.035\,$\pm$\,0.005 & 63\,$\pm$\,44 & 64\,$\pm$\,10 & 63\,$\pm$\,44 & &3.5\,$\pm$\,1.4&\\ \hline
   0.075 & 1.9\,$\pm$\,0.2 & 5.6\,$\pm$\,0.8 & -\,0.026\,$\pm$\,0.005 & 49\,$\pm$\,37 &51\,$\pm$\,10 & 49\,$\pm$\,37 & &3.8\,$\pm$\,1.7 &\\ \hline
   0.063& 1.8\,$\pm$\,0.3 & 6.0\,$\pm$\,0.5 & -\,0.022\,$\pm$\,0.005 & 40\,$\pm$\,21& 43\,$\pm$\,10& 40\,$\pm$\,21& &3.7\,$\pm$\,1.1 &\\ \hline
   0.05& 1.7\,$\pm$\,0.2 & 5.7\,$\pm$\,0.6 & -\,0.023\,$\pm$\,0.005 & 39\,$\pm$\,25 & 41\,$\pm$\,10 & 39 \,$\pm$\,25 & &3.5\,$\pm$\,1.3  &\\ \hline
   0.03& 2.0\,$\pm$\,0.3 & 6.1\,$\pm$\,0.4 & -\,0.016\,$\pm$\,0.005 & 32\,$\pm$\,15 & 32\,$\pm$\,10 & 32\,$\pm$\,15 & &4.02\,$\pm$\,0.85 & \\
    \hline
    \end{tabular}}
    \begin{minipage}{\textwidth}
    \caption{Data table for ARPES kink analysis for OPT LSCO ( T = 20 K ) \cite{Zhou2003} in Fig.~\ref{LSCOEDC}. We were unable to reliably estimate  
    $\Gamma_0 $ here due to the lack of data at high temperature, and hence set it at zero. The uncertainties for measured values were given by half of the instrumental resolution (10 meV, $\sim$0.005\,$\AA^{-1}$). The uncertainties for the calculated values were determined by error propagation.}
    \label{table2}
    \end{minipage}
    \end{center}
\end{table}
\FloatBarrier
Our analysis  becomes unreliable as lower doping level $\text{x} < 0.075$ in Panels (h) to (j) in Fig.~\ref{LSCOEDC}, 
where the dispersion kink is no longer a simple bending kink,  an extra curving tendency begins to appear. 
To put this in context, recall that the line shape of LSCO becomes extremely broad at small x \cite{Yoshida2007}, and so the peak position of the spectral function becomes more uncertain than at higher energy.
We should point out that in \figdisp{LSCOEDC} Panel (k) the spectral function has been shifted  to right by 4 meV for a better fit. This shifting  is  consistent with our argument  that the Fermi momentum determination has a possible small error of  in order  0.006\,$\AA^{-1}$, arising from  the $\hat{k}$ dependent  caparison factor,  and hence the peak position has an uncertainty   $V_L \times .006 \sim10$ meV.

\section{Bi2201 Laser ARPES data}
In this section, we present our  analysis of the high resolution laser ARPES data of the single layered compounds Bi2201, at various different doping levels taken from a recent study in \refdisp{Ying2013}.
In earlier studies of this compound using synchrotron emitted high energy photons,  as also LSCO \cite{Lanzara2001},  the ARPES kinks  were observed to have only a weak  temperature dependence \cite{Sato2003}.
However, the new high resolution laser ARPES data  enables us to observe clear and significant temperature dependence of the ARPES kinks; it is comparable to that of the double layered Bi2212 compounds. In fact we find that the new data of Bi2201 compounds in \refdisp{Ying2013} seems to provide  a textbook example of our  ECFL kink analysis.

\begin{widetext}
\begin{figure}[htp]
\includegraphics[width= 0.9\textwidth]{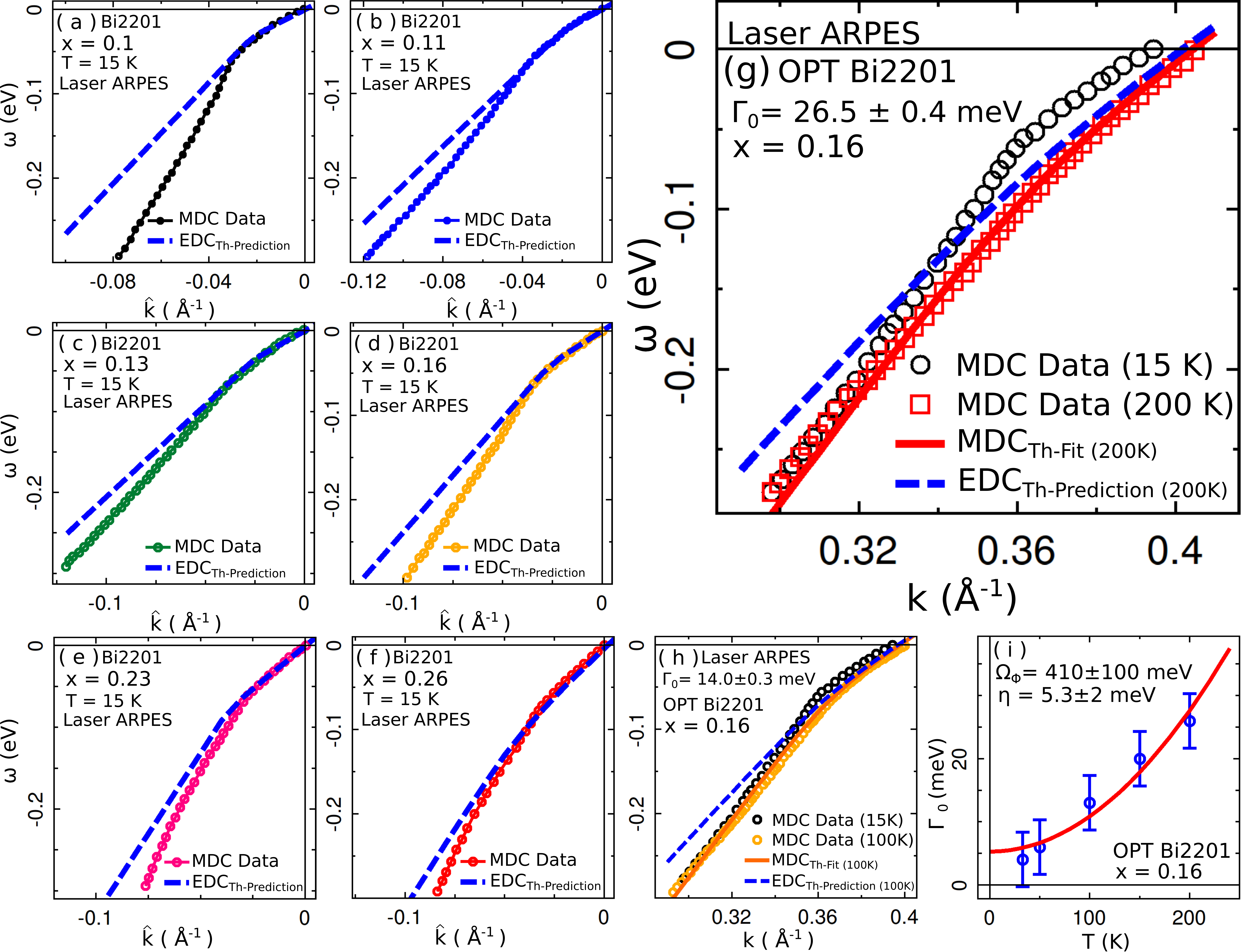}
\begin{minipage}{\textwidth}
\caption{\footnotesize ARPES kink analysis for laser ARPES data of Bi2201 at various different doping levels in \refdisp{Ying2013}. {\bf Panels:(a) to (f)} we predict EDC dispersions (blue dashed lines) using \disp{EDC} for various different doping levels of Bi2201 laser ARPES data. {\bf Panel:(g) and (h)} we predict high temperature EDC (blue dashed lines) dispersions (g) 200K and (h) 100K for laser ARPES data of OPT Bi2201 (panel (a) in Fig.4 in \refdisp{Ying2013}), and show the MDC dispersion fits for two temperature also. We estimate $\Gamma_0$ from measuring the difference between the ideal kink energy and the MDC kink energy. In order  to compare with experiments, the x-axis representation in (g) and (h) are given by the physical $k$ (rather than the momentum difference $\hat{k}$).  In panel (g), the MDC dispersion fit of 200 K vanishes  at  $k = 0.404\,\pm\,0.002\,\AA^{-1}$,   very close to the measured $k = 0.405\,\pm\,0.002\,\AA^{-1}$ of the MDC dispersion data at 200 K. Similarly in panel (h) the MDC dispersion fit at 100 K vanishes  at $k = 0.398\,\pm\,0.002\,\AA^{-1}$, close to the measured $k = 0.4\,\pm\,0.002\AA^{-1}$ of the MDC dispersion data at 100 K. Note that the true fermi momentum as estimated from the low T (15 K) data is $k = 0.394\,\pm\,0.002\,\AA^{-1}$, so that the deviations are  bigger than the momentum resolution $\Delta k\sim 0.004\,\AA^{-1} $. 
 {\bf Panel:(i)} we plot the temperature dependence of $\Gamma_0$ in panel (a) in Fig.4 in \refdisp{Ying2013}. Here, the temperature dependence data of $\Gamma_0$ is fitted with \disp{rdef}, and $\eta$ is determined 5.3 $\pm$ 2 meV and $\Omega_\Phi$ = 410 $\pm$ 100 meV. 
}  
\label{laserBi2201}
\end{minipage}
\end{figure}
\end{widetext}

In Table(\ref{table3}) we list the kink parameters corresponding to different doping levels of Bi2201 and tabulate the kink parameters. The entries are in correspondence to the panels in \figdisp{laserBi2201}. In  \figdisp{laserBi2201} panels (a) to (f),  we depict the  measured  MDC dispersion and the predicted  EDC dispersions  at  different doping levels. The latter are found  from \disp{EDC} using the  variables in Table~(\ref{table3}).  Panels (g) and (h) of OPT Bi2201 are especially interesting. Combining  the low $T=15$K dispersion data  and the finite T value of $\Gamma_0$, found from the depression of the kink energy $E^{MDC}_{kink} = E^{ideal}_{kink}- \Gamma_0 \sqrt{\frac{r}{2-r}}$, we can   reconstruct   the entire MDC dispersion at a finite  T. This may be    compared  with the measured finite T MDC data, thus  checking the validity of the formalism. This exercise is carried out at T=200K in Panel (g) and T=100 K in panel (h), where we find a remarkably good fit in all details. In panels (g,h) we show the actual momentum (rather than $\hat{k}$) to facilitate a comparison with data. Panel (g) especially clearly shows that $E(\hat{k})$ vanishes at a  $\hat{k}$ that is different  from 0. The shift corresponds to  $\sim 0.01 \AA^{-1}$.
We have commented above that this  apparent expansion of the Fermi surface with T  is  due to the non trivial physics underlying \disp{MDC} lying beyond the simple minded FLT.

Panel (i) in \figdisp{laserBi2201} plots the temperature dependence of $\Gamma_0$ in panel (a) in Fig.(4) in \refdisp{Ying2013}. The measured $\Gamma_0$ curve is fitted with \disp{rdef}, and we estimate $\eta = 5.3 \pm 2$ meV and $\Omega_\Phi = 410 \pm 100$ meV.

\begin{table}[htp]
\begin{center}
 \resizebox{\textwidth}{!}{\begin{tabular}{|c|c|c|c||c c|c c||c c|}  
   \hline
   \multicolumn{6}{|c|}{MDCs} &\multicolumn{4}{c |}{EDCs}\\
   \hline
   \multicolumn{4}{| c ||}{Bi2201 laser ARPES data} & \multicolumn{2}{c |}{$E^{MDC}_{kink}$(meV)}&\multicolumn{2}{c ||}{$E^{EDC}_{kink}$(meV)} &\multicolumn{2}{c |}{$V_H^*$ (eV\,$\AA$)}\\
   \hline
 x\,(\,{\footnotesize doping level}\,) & $V_L$ (eV\,$\AA$) & $V_H$ (eV\,$\AA$) & $k_{kink}$ ($\AA^{-1}$) & Calculated & Measured & Calculated & Measured &Calculated & Measured\\ 
 \hline
      0.1 & 1.47\,$\pm$\,0.12 & 4.7\,$\pm$\,0.3 & -\,0.022\,$\pm$\,0.002 & 32\,$\pm$\,3 & 37\,$\pm$\,0.5 &32\,$\pm$\,6 & & 3.0\,$\pm$\,0.3&\\ \hline
     0.11 & 1.34\,$\pm$\,0.06 & 2.78\,$\pm$\,0.06 & -\,0.021\,$\pm$\,0.002 & 28\,$\pm$\,1 & 28\,$\pm$\,0.5 &28\,$\pm$\,4  & & 2.28\,$\pm$\,0.12&\\ \hline
     0.13 & 1.37\,$\pm$\,0.07 & 2.71\,$\pm$\,0.18 & -\,0.025\,$\pm$\,0.002 & 38\,$\pm$\,3 & 39\,$\pm$\,0.5& 37\,$\pm$\,5 & & 2.27\,$\pm$\,0.17&\\ \hline
     0.16 & 1.5\,$\pm$\,0.1 &3.5\,$\pm$\,0.2 & -\,0.026\,$\pm$\,0.002 & 39\,$\pm$\,3 & 43\,$\pm$\,0.5& 39\,$\pm$\,6 & &2.7\,$\pm$\,0.2&\\ \hline
     0.23 & 2.1\,$\pm$\,0.11  & 5.4\,$\pm$\,0.3 & -\,0.036\,$\pm$\,0.002 & 98\,$\pm$\,6 &97\,$\pm$\,0.5 &89\,$\pm$\,10 &  & 3.9\,$\pm$\,0.3&\\ \hline
     0.26 & 2.17\,$\pm$\,0.16  & 4.8\,$\pm$\,0.4 & -\,0.045\,$\pm$\,0.002 & 123\,$\pm$\,11 &122\,$\pm$\,0.5 &114\,$\pm$\,18 &  & 3.8\,$\pm$\,0.4&\\ \hline\hline
     0.16\,(\,200\,K\,) & 1.61\,$\pm$\,0.18  & 3.5\,$\pm$\,0.3 & 0.364\,$\pm$\,0.002 & 87\,$\pm$\,11 & 89\,$\pm$\,0.5 &75\,$\pm$\,11 &  & 2.8\,$\pm$\,0.4&\\ \hline
     0.16\,(\,100\,K\,) & 1.61\,$\pm$\,0.18  & 3.5\,$\pm$\,0.3 & 0.364\,$\pm$\,0.002 & 69\,$\pm$\,11 & 70\,$\pm$\,0.5 &62\,$\pm$\,11 &  & 2.8\,$\pm$\,0.4&\\
    \hline
    \end{tabular}}
    \begin{minipage}{\textwidth}
    \caption{Parameter table for ARPES kink analysis for laser ARPES data of Bi2201 at various different doping levels \cite{Ying2013} in Fig.\ref{laserBi2201}. From 0.1 $<$ x $<$ 0.16, we measured $\Gamma_0 \sim 0 $. For x = 0.23 and 0.26, we measured $\Gamma_0 \lesssim$ 17 meV. For x = 0.16 data, we report variables for high temperature kinks data 200 K (g) and 100 K (h) in \figdisp{laserBi2201}, and $\Gamma_0$ values for 200 K and 100 K data are in corresponding panels (g) and (h) in \figdisp{laserBi2201}. The uncertainties for the calculated parameters were determined by error propagation, and 
    the uncertainties for the experimental parameters were given by half of the instrumental resolution.}
    \label{table3}
    \end{minipage}
  \end{center}
 \end{table}
 \FloatBarrier

\section{conclusion}

The main goal of this work is to understand the physical origin of kinks in the dispersion seen in ARPES studies of a wide class of systems. For this purpose we have listed fifteen systems of topical interest where ARPES kink data is available. Our focus is on the nodal direction data, since the largest volume  is available here.  We have devised a useful protocol to extract kink parameters from data, where the  asymptotic tangents of the kink  are used. Using this protocol we have analyzed in detail three families of systems, two synchrotron and one laser ARPES data of cuprate superconductors. The main parameters of the kinks are the energy, momentum and the dispersion velocities in EDC and MDC scans, these provide a  quantitative  data set  for testing various theoretical proposals for explaining kinks. 

We have outlined two competing theories for the origin of kinks, and highlighted their distinctive predictions. One is the electron-Boson model, where an Einstein mode of either spin or charge origin
couples to the electrons, resulting in   a momentum independent self energy.  This theory  gives rise to   kinks in the electron dispersion. The other theory is the strong or extreme correlation theory, where the interactions lead to a momentum dependent self energy in two dimensions. This theory also gives rise to kinks in  the electron dispersion.

 The predictions of the two theories differ  significantly and in experimentally testable ways. The Boson-mode theory gives rise to kinks located  {\em at} the energy of the localized mode. 
For the kinks, the Boson-mode theory  predicts \cite{supplement}: 
(1) a momentum independent peak in the spectral function at the kink energy when $\hat{k}< \hat{k}_{kink}$, 
(2) a jump in the EDC dispersion at the kink energy but not the MDC dispersion and 
(3) the EDC and MDC velocities are identical both before and after the kink is crossed. 

The extremely strong correlation theory also gives rise to kinks in dispersion, these originate from the momentum dependence of the self energy \cite{supplement}. A simple  low energy and momentum expansion of the ECFL  theory gives inter-relations between observed features of the kinks. It predicts
(1) a kink at an emergent low energy scale originating from Gutzwiller correlations 
(2) no jump in the EDC dispersion and 
(3) the EDC velocity is determined by the MDC velocities through $V^*_H= \frac{3 V_H -V_L}{V_H+V_L} \times V_L$. 
It is remarkable that a knowledge of the MDC dispersion suffices to predict the EDC dispersion, and
the parameters obtained from the MDC dispersion enable us to reconstruct the spectral function at low momentum and energy, in both MDC and EDC scans.

It is thus clear that EDC dispersions hold the key to distinguishing between the two competing theories.  EDC dispersion data  is sparse but exists,  the  work on  OPT Bi2212 from \refdisp{Kaminski2001} shown in   \figdisp{fig2},  presents   both EDC and MDC dispersions  at 115 K. Its resolution is presumably not optimal, since it was an early  experiment. Nevertheless we  can use it to make a first pass at comparing  the two theories. This data set plotted in \figdisp{fig2}  shows that the EDC dispersion is continuous, i.e. has no jump.  Further the EDC higher velocity $V_H^*$  is close to that predicted by the ECFL analysis. The measured spectral function in EDC,  overlooking the  noise,  seem not  to have any  immovable feature at $E_{kink}$. Thus all three characteristics noted above appear to be consistent with the ECFL predictions rather than the Bosonic mode theory predictions.
It  is  roughly  fit by the low energy parameterized curves as well, where the MDC is seen to be more symmetric than the EDC cuts.

As noted in Table~(\ref{survey}) the above case OPT Bi2212 is particularly interesting. Low energy Bosonic modes have been observed in neutron scattering \cite{Fong1999, He2001}, and in  momentum resolved electron energy loss experiments \cite{Vig2015}.  In \refdisp{Vig2015}  an MDC dispersion is presented using  parameters taken  from the  Bosonic data. This leads to a rather detailed  model,  and is shown  to provide a reasonable fit to the MDC dispersion and the observed kink, but the important EDC dispersion is not displayed.

While we focussed attention on dispersion kinks in  the nodal direction in the present work,  the ECFL theory is also valid for other directions, it has a momentum dependence in the self energy both normal to the Fermi surface and also along the tangent.  The  ECFL theory applied to the  d-wave superconducting state in the $t$-$J$ model  is expected to lead to further  interesting results in the future.  For now we note  that the observed  nodal direction spectra are essentially unchanged at $T_c$, which makes the nodal direction  particularly interesting. 

In conclusion, we have presented a  current summary of the physics of the kinks in dispersion of cuprate high Tc superconductors. We believe that  there is  urgent  need for  further high resolution  EDC data, and also  T dependent scans to explore the rounding of kinks.  Using such data  one should be able to check the predictions of the  theory more thoroughly, and thereby obtain definitive understanding of the origin of low energy ARPES kinks of strongly correlated matter.

\section{Acknowledgements}
We thank Antoine Georges  and Jason Hancock  for stimulating discussions. The work at UCSC was supported by the U.S. Department of Energy (DOE), Office of Science, Basic Energy Sciences (BES) under Award \# DE-FG02-06ER46319.

\end{document}


\title{ Supplementary Information:\\  Origin of  Kinks in  Energy Dispersion  of Strongly  Correlated Matter}
\author{ Kazue Matsuyama}
\affiliation{Physics Department, University of California, Santa Cruz, CA 95064 }
\author{Edward Perepelitsky }
\affiliation{Centre de Physique Th\'eorique, \'Ecole Polytechnique, CNRS, Universit\'e Paris-Saclay, 91128 Palaiseau, France}
\affiliation{Coll\`ege de France, 11 place Marcelin Berthelot, 75005 Paris, France}
\author{B Sriram Shastry}
\affiliation{Physics Department, University of California, Santa Cruz, CA 95064 }
\date{\today}
\maketitle
\makeatletter
\renewcommand\@biblabel[1]{[SI-#1]}
\makeatother

 In this supplemental note we provide (I) some details of the doping dependence of the fit parameters
(II) detailed predictions  of the electron-Boson coupling model  for kinks and (III)  detailed predictions  of the extremely  strong correlation theory for kinks. In the main paper, we have discussed alternate  mechanisms  for generating  the low-energy kink observed in ARPES. Although both mechanisms are capable of generating similar MDC dispersions, they  produce EDCs and  EDC dispersions which are distinct from one another in several clearly identifiable ways. These differences, detailed below, can be used to distinguish between the two mechanisms using ARPES, especially as higher resolution data becomes available in the future.

\subsection{Fixing the parameters}
The independent parameters in the ECFL expressions  for  the kink can be taken as  $V_H,V_L,\kk_{kink}$ and $\Gamma_0$.  These can  be fixed with four measurements as we indicate below. 
While the first three can be measured with precision, the variable $\Gamma_0$ depends on the temperature and  is also quite sensitive to the various  experimental conditions including the incident photon energy, thus making it less precisely known than the others; we will perforce be content with rough estimates of this variable.  
The remaining  parameters can be calculated using equation (MS-2) and equation (MS-5) etc.
  As mentioned above, the theory is overdetermined, 
in terms of these four parameters, the theory predicts a number of  other quantities: 
a) the dispersion curves for both EDCs and MDCs,
b) the location of both EDC and MDC kinks at finite temperature, and
c) the spectral functions near the Fermi level (\,up to roughly the kink energy). 
Below we present an analysis of the ARPES data of Bi2212, LSCO and Bi2201 taken from literature, where we give the details of the fits and the predicted EDC velocities for future experiments.

 The asymptotic velocities $V_H,V_L$  determine the ratio $r$ from equation (MS-2). The energy $\Delta_0$  and the { ideal kink energy} are determined from equations~(MS-5, \ref{ideal}).
As discussed in Fig.~1 $ E^{MDC}_{kink}$ is found by measuring the dispersion at the kink wave vector $E(\hat{k}_{kink})$,
and similarly the EDC kink energy    $ E^{EDC}_{kink}$ is found from $E^*(\hat{k}_{kink})$.
For understanding the finite temperature data, the theory provides   temperature dependent correction terms for the two spectra,  determined  by the parameter $\Gamma_0$,
 \beq 
 E^{EDC}_{kink}& =&  E^{ideal}_{kink} - \Gamma_0, \label{EDC-kink}\\
 E^{MDC}_{kink} &=& E^{ideal}_{kink}- \Gamma_0 \sqrt{\frac{r}{2-r}}. \label{MDC-kink}
 \eeq
Since  $\Gamma_0$ determines the non-zero T (or $\eta$) correction, we estimate  from the difference between low and high temperature MDC dispersion curves
\beq
\Gamma_0 = \Delta E_{kink}=\sqrt{\frac{2-r}{r}} \left( E^{ideal}_{kink} - E^{MDC}_{kink} \right). \label{gamma0}
\eeq
Clearly uncertainties in $\Gamma_0$ are governed by those in the MDC dispersion at the kink momentum. 

As noted in Fig.~1, the ECFL   theory predicts a kink, rather than a jump in the EDC spectrum, quite analogous to that in the MDC dispersion, but with a different velocity on the steeper side, i.e. $V^*_H \neq V_H$. In fact the theory provides an experimentally testable  expression  relating the two,  $V_H^*$  is expressed quite simply  in terms of measurable experimental  variables,
\beq
V^*_H= \frac{3 V_H -V_L}{V_H+V_L} \times V_L.  \label{Vh-EDC}
\eeq
As mentioned in the introduction the Boson-mode coupled   theories predict  a jump in the EDC spectrum at the kink energy.  The velocity beyond the jump is the same in EDC and MDC, i.e. $V_H^* =V_H$, in contrast to \disp{Vh-EDC}.
This velocity is reported in only a few cases, and provides a ready test of the ECFL theory.

The theory also predicts $V_L=V_L^*$, which is satisfied by inspection in all reported cases and is common to the Boson-mode theory. We  use this protocol to analyze the experiments on three well studied families of high $T_c$ materials next. 

\subsection{Fit parameters}

\begin{widetext}
 \begin{figure}[h]
\includegraphics[width = 0.85\columnwidth]{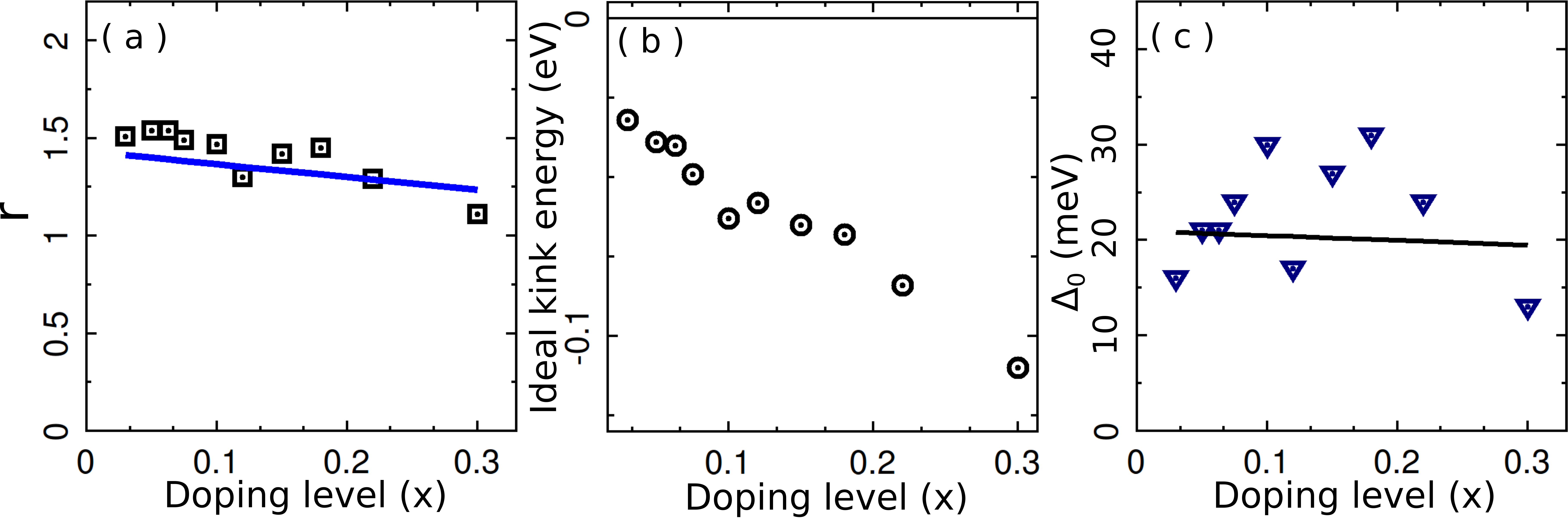}
\caption{ ( a ) The ratio of low and high velocities, r, as a function of doping levels, 
and ( b ) ideal kink energy, ( c ) ECFL energy parameter $\Delta_0$ as a function of doping levels for LSCO data in the main text.} 
\label{Ratio}
\end{figure}
\end{widetext}

{\bf (\RomanNumeralCaps{1}) $\Delta_0$ for LSCO data in the main text}
For the  LSCO data discussed in the main text, we quoted the  ECFL theory  parameters, velocity ratio r, the ideal kink energy $E^{ideal}_{kink}$  and the small energy parameter  $\Delta_0$, in Eqs~(1,6,4) (see also  \disp{ideal}).
In Fig.~\ref{Ratio}, we display the doping dependence of these parameter $x=1-n$. The size of the data point represent the uncertainty for each data points. 
While  r and $\Delta_0$ stay almost  constant,  the ideal kink energy  decreases linearly with increasing x.  

\section{(II)  Electron-Boson coupling theory of kinks}

The electron Boson mechanism suggested in \refdisp{Cuk2005} and others  \cite{He2013,Vig2015}, is the coupling of the electrons to Bosonic modes (such as phonons), located  at the kink energy.  To illustrate the basic idea,  we first consider free electrons coupled to an Einstein phonon mode of energy $\omega_0=.08  \ \text{eV}$ \cite{He2013,Vig2015}, with coupling constant $g$. In this case, the spectral function is expressed in terms of a momentum independent self-energy $\Sigma(\omega)$, as

\beq
A(\vk,\omega) = -\frac{1}{\pi} \frac{\Im m \Sigma(\omega)}{(\omega-\xi_k- \Re  e \Sigma(\omega))^2 + (\Im m \Sigma(\omega))^2},\nn\\
\label{A_ph}
\eeq
where $\xi_k\equiv \varepsilon_k-\mu$, $\varepsilon_k$ is the bare dispersion, and $\mu$ is the chemical potential. The real and imaginary parts of the self-energy due to the electron-phonon interactions are given by the well known formulas: \cite{Migdal1958,Engelsberg1963}
\beq
\Im m \Sigma(\omega) &=& - \pi g^2 \sum_{\pm} N(\omega+\mu\pm\omega_0)\times\nn\\
&&\left[f^{\mp}(\omega\pm\omega_0)+n(\omega_0)\right],\nn\\
\Re e \Sigma(\omega) &=& -\frac{1}{\pi} \int d\nu \frac{\Im m \Sigma(\nu)}{\omega-\nu},
\label{Sigmaeph}
\eeq
where $f^-(\nu) \equiv f(\nu)$, $f^+(\nu) \equiv \bar{f}(\nu)\equiv 1- f(\nu)$, $f(\nu)$ and $n(\nu)$ are the Fermi and Bose distribution functions respectively, and $N(E)\equiv \frac{1}{N_s}\sum_k \delta(E-\varepsilon_k)$ is the local density of states for the free electrons. Since the relevant frequency range for the self-energy is $|\omega| \sim \omega_0$, and $\omega_0\ll W$, where $W$ is the bandwidth, we neglect the frequency dependence in the density of states, i.e. $N(\omega+\mu\pm\omega_0)\approx N(\mu)\approx N(\varepsilon_f)$, where $\varepsilon_f$ is the Fermi energy. Furthermore, the strength of the electron-phonon coupling is given by the dimensionless parameter \cite{Gunnarson} $ \lambda\equiv\frac{2 N(\varepsilon_f)g^2}{\omega_0}$. Therefore, the imaginary part of the self-energy is expressed directly in terms of $\lambda$ as

\beq
\Im m \Sigma(\omega) &=& - \frac{\pi\lambda\omega_0}{2}  \sum_{\pm}\left[f^{\mp}(\omega\pm\omega_0)+n(\omega_0)\right].\nn\\
\label{ImSigmaeph}
\eeq
\begin{figure}[htp]
\begin{center}
\includegraphics[width=0.33\columnwidth]{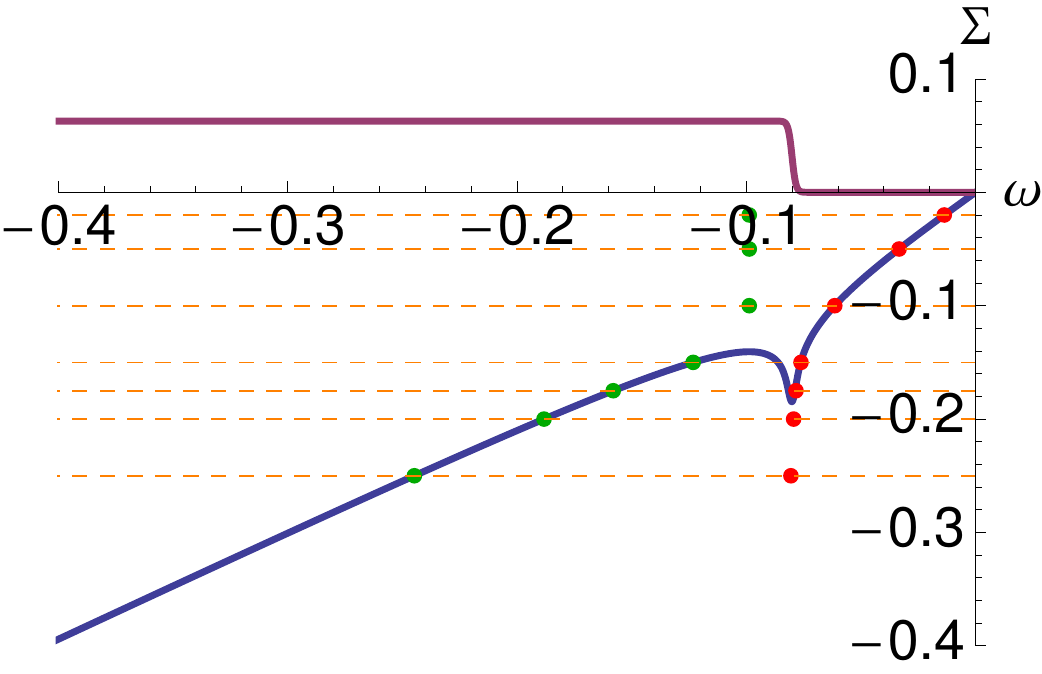}
\includegraphics[width=0.32\columnwidth]{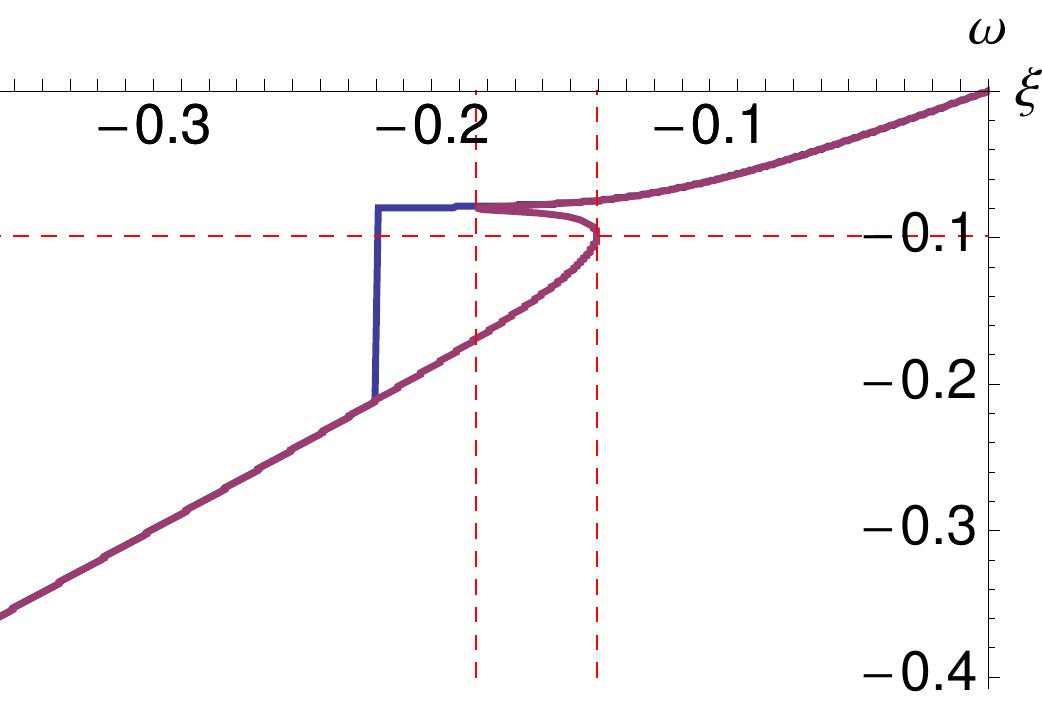}
\includegraphics[width=0.33\columnwidth]{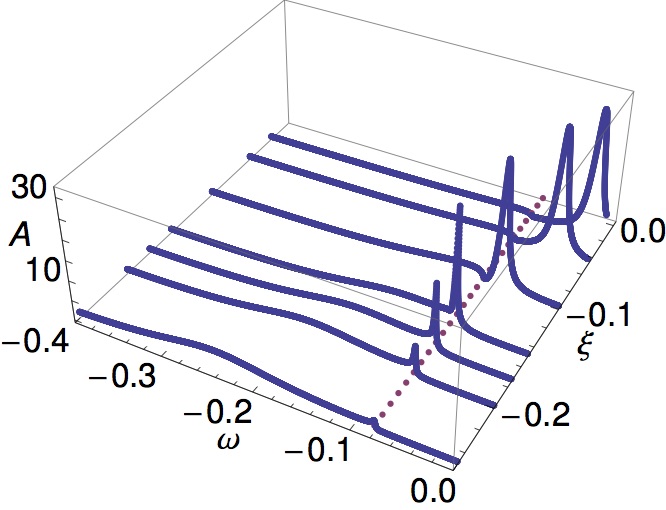}
\caption{Results for free electrons coupled to an Einstein phonon mode of frequency $\omega_0=.08  \ \text{eV}$, with coupling strength $\lambda=0.5$, at $T= 10 \ \text {K}$. {\bf Right } panel: The EDCs at several representative momenta, the variable $\xi=v_f(k-k_F)=(1+\lambda)V_L(k-k_F)$ here and in later figures. The dashed line indicates the phonon frequency, $\omega=-\omega_0$. Each EDC has two well-defined features, a peak followed by a hump (separated by a sharp dip for low momentum EDCs). {\bf Middle} panel: The MDC dispersion (magenta) has no jump while  the EDC dispersion (blue) shows a jump. The two vertical dashed lines partition momentum space into three regions. The horizontal dashed line indicates the location of the hump in the EDCs in the first (low-momentum) region. In the first two regions, the EDC dispersion follows the MDC dispersion (closest to zero frequency), while in the third (high momentum) region, it stays pinned to the phonon frequency over a large range of momentum, until it discontinuously jumps back down to the MDC dispersion. Note that $V_H=V_H^*$. {\bf Left} panel: $\omega-\Re e \Sigma(\omega) $ and $-\Im m \Sigma(\omega)$ vs. $\omega$. The horizontal dashed lines indicate the momenta associated with the corresponding EDCs in the right panel. The red dots indicate the locations of the peaks, and the green dots indicate the locations of the humps, as determined directly from each EDC.} 
\label{EDC_MDC_FG} 
\end{center}
\end{figure}
\FloatBarrier
We initially  choose a typical intermediate strength value of $\lambda=0.5$. We also add a small broadening $\eta = .01 \ \text{eV}$ to the imaginary part of the self-energy. In \figdisp{EDC_MDC_FG}, we display $\omega-\Re e \Sigma(\omega) $ and $-\Im m \Sigma(\omega)$ vs. $\omega$ (left panel), the EDC and MDC dispersions (middle panel), as well as the EDCs at several representative momenta (right panel) at $T= 10 \ \text {K}$. The EDC and MDC dispersions as well as the EDCs can be understood directly from the real and imaginary parts of the self-energy using \disp{A_ph}. From \disp{A_ph}, the the MDC at fixed $\omega$ is a Lorentzian of width $-\Im m \Sigma (\omega)$ and peak position $ \xi^*(\omega)=\omega- \Re e \Sigma(\omega)$ \cite{Cuk2005}. Therefore, the MDC dispersion is obtained by inverting $\xi^*(\omega)$ to obtain $E(\xi)$. Since $\omega- \Re e \Sigma(\omega)$ is not one-to-one, $E(\xi)$ is a multi-valued function.

To understand the EDC dispersion, we first examine the EDC curves in the right panel of \figdisp{EDC_MDC_FG}. The momentum $\xi$ associated with each curve is given by the location of the corresponding horizontal dashed line along the vertical axis in the left panel. The EDC at each momentum has two distinguishable features, a peak followed by a hump. In the left panel, the red and green dots indicate the location of the peak and hump, respectively, at each momentum, as determined directly from the EDC.

We partition the EDCs into three distinct momentum regions, $|\xi|<|\xi_1|$, $|\xi_1|<|\xi|<|\xi_2|$, and $|\xi|>|\xi_2|$, where the momenta $\xi_1$ and $\xi_2$ (the low-energy kink momentum) are denoted by the dashed vertical lines in the middle panel of \figdisp{EDC_MDC_FG}. In the first region, $|\xi|<|\xi_1|$, the peak location, $E^*_p$, disperses according to the equation $\xi=E^*_p- \Re e \Sigma(E^*_p)$, while the hump location, $E^*_h$, remains at a fixed frequency, displayed by the horizontal dashed line in the middle panel. In addition, there is a sharp dip between the peak and the hump which is pinned to the phonon frequency, $-\omega_0$. Since $\Im m \Sigma(E^*_p)$ is constant throughout this region, the height of the peak does not change. On the other hand, since $|E^*_h- \xi-\Re e \Sigma(E^*_h)|$ decreases as $|\xi|$ is increased (and of course $\Im m \Sigma(E^*_h)$ is constant), the hump height grows as $|\xi|$ approaches $|\xi_1|$. Nevertheless, since the peak height remains greater than the hump height throughout this region (as will be shown below), the EDC dispersion is given by $E^*=E^*_p$.

In the second region, $|\xi_1|<|\xi|<|\xi_2|$, both $E^*_p$ and $E^*_h$ disperse according to the equation $\xi=E^*_{p,h}- \Re e \Sigma(E^*_{p,h})$, $E^*_p$ being the root closest to, and $E^*_h$ being the root farthest from, zero frequency. Since  $\Im m \Sigma(E^*_p)$ continues to remain constant and has the same value as in the first region, so does the height of the peak. Moreover, since $\Im m \Sigma(E^*_h)$ remains constant as well, the height of the hump remains the one which it reached at $\xi=\xi_1$. Finally, since $|\Im m \Sigma(E^*_h)|>|\Im m \Sigma(E^*_p)|$, the peak height is greater than the hump height, and therefore $E^*=E^*_p$.

In the third region, $|\xi|>|\xi_2|$, $E^*_p$ is pinned to the phonon frequency $-\omega_0$, while $E^*_h$ continues to disperse according to the equation $\xi=E^*_{h}- \Re e \Sigma(E^*_{h})$. Since $\Im m \Sigma(E^*_h)$ continues to have the same value as in the second region, so does the height of the hump. Meanwhile, the peak height decreases, since $|E^*_p- \xi-\Re e \Sigma(E^*_p)|$ increases as $|\xi|$ is increased. Although initially $E^*=E^*_p=-\omega_0$, eventually, after $|\xi|$ has been sufficiently increased, the peak height falls below the hump height, and $E^*=E^*_h$. Accordingly, in the middle panel, we see that in first two regions, the EDC dispersion follows the MDC dispersion, $E^*=E$ (closest to zero frequency). However, in the third region, $E^*$ stays fixed at $-\omega_0$, until at sufficiently high momentum, it jumps back down to the MDC dispersion. Since the MDC and EDC dispersions coincide for large momentum, the velocities $V_H$ and $V_H^*$ are equal. We take these three features, a discontinuous jump in the EDC dispersion, a peak pinned to the phonon frequency in the EDC over a prolonged range of momentum, and the equality $V_H=V_H^*$, to be signatures of electron-Boson coupling in ARPES experiments. Similar calculations to the one above can be found in \cite{Cuk2005,He2013}, with analogous results.

\begin{figure}[htp]
\begin{center}
\includegraphics[width=0.33\columnwidth]{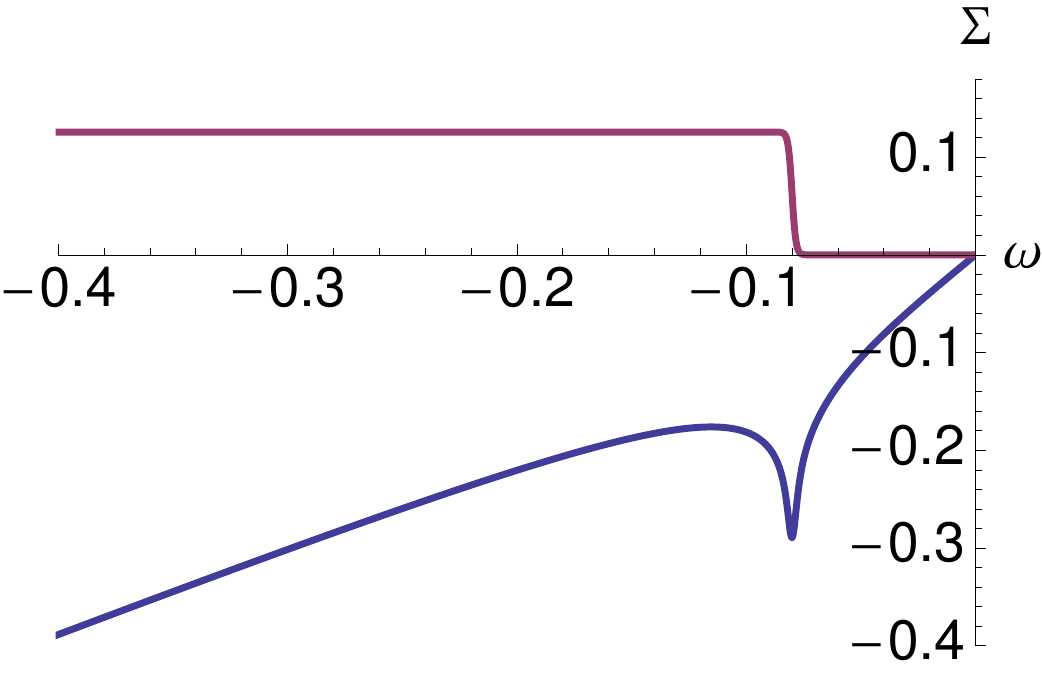}
\includegraphics[width=0.32\columnwidth]{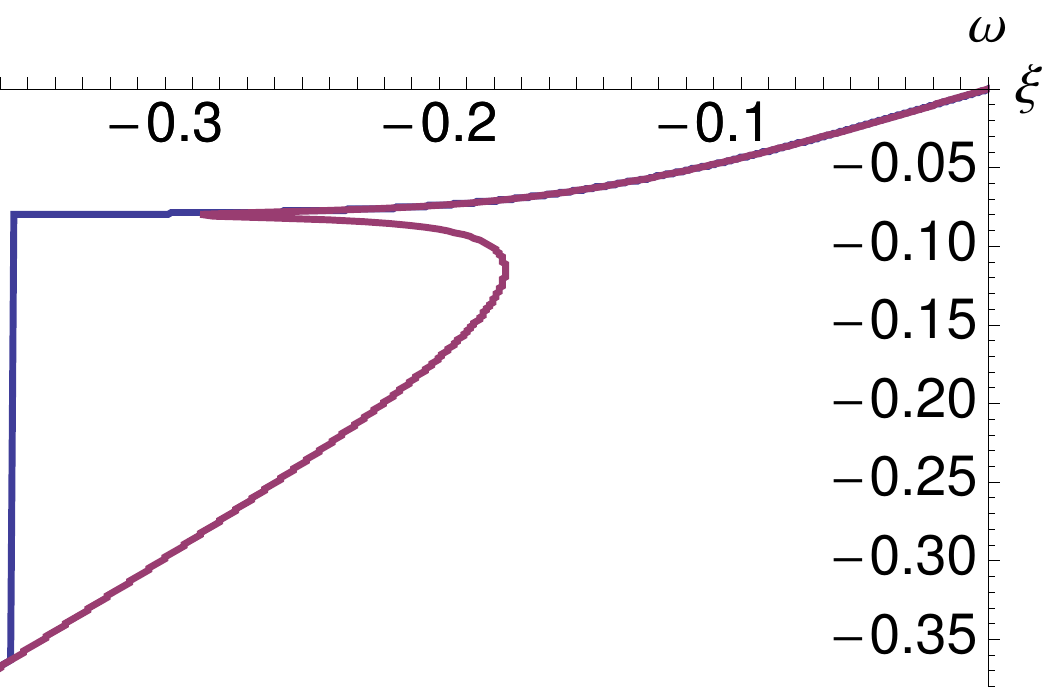}
\includegraphics[width=0.33\columnwidth]{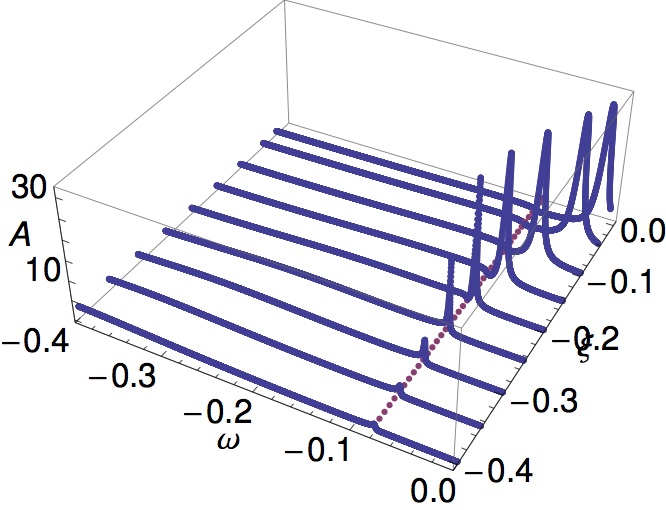}
\caption{To explore the effects of raising $\lambda$, we set $\lambda=1$ while leaving all other parameters unchanged from \figdisp{EDC_MDC_FG}. As a result, the kink momentum in the MDC dispersion becomes bigger, the hump in the EDCs is suppressed, the EDC dispersion stays pinned to the phonon frequency over a larger range of momentum, and the magnitude of the jump in the EDC dispersion grows.} 
\label{EDC_MDC_lambda_1} 
\end{center}
\end{figure}
\FloatBarrier
To examine the effects of raising $\lambda$, we set $\lambda=1$ leaving all other parameters unchanged, and plot the corresponding results in \figdisp{EDC_MDC_lambda_1}. This causes several noticeable changes to the results in \figdisp{EDC_MDC_FG}. 1) The kink in the real part of the self-energy becomes sharper, which leads to a larger kink momentum, $\xi_2$, in the MDC dispersion. 2) $-\Im m \Sigma(E^*_h)$ becomes bigger, causing the height of the hump to go down. 3) As a direct consequence of 2), the range over which the EDC dispersion stays pinned to the phonon frequency becomes more prolonged in momentum space, and therefore the magnitude of the jump in the EDC dispersion also becomes bigger.

\begin{figure}[htp]
\begin{center}
\includegraphics[width=0.33\columnwidth]{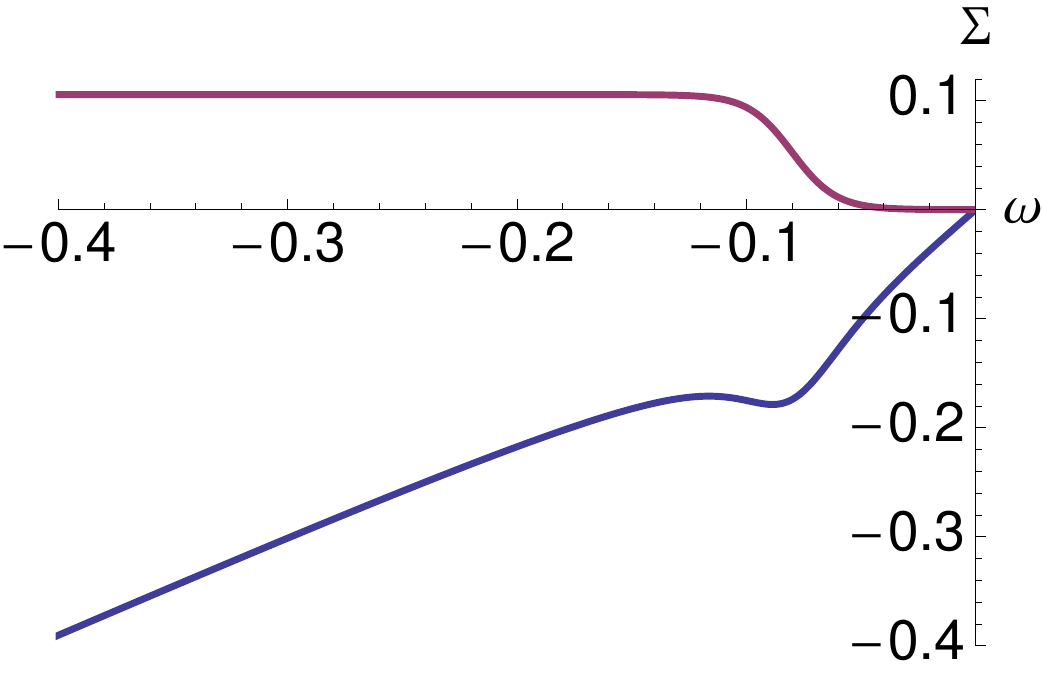}
\includegraphics[width=0.32\columnwidth]{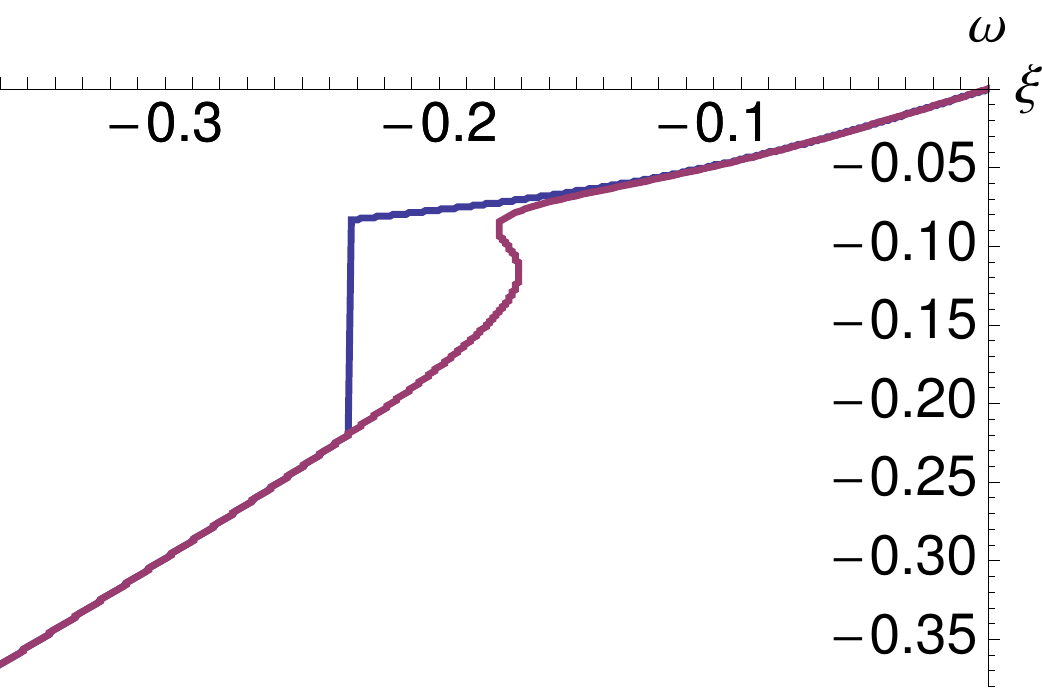}
\includegraphics[width=0.33\columnwidth]{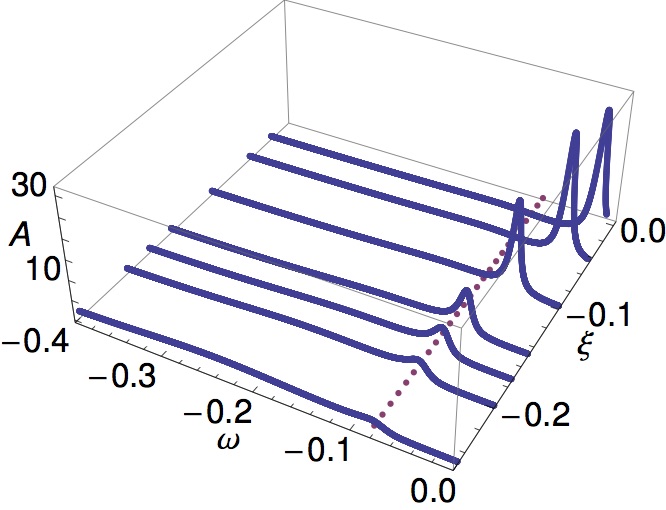}
\caption{We now use the experimentally relevant values of $\lambda = 0.84$ and $T=115 \ \text {K}$. The curves retain the same qualitative features as in \figdisp{EDC_MDC_FG}, which are less sharp in the present case due to the higher value of $T$. } 
\label{EDC_MDC_FG_T115} 
\end{center}
\end{figure}

Setting $T\to0$ in \disp{ImSigmaeph}, and plugging it into \disp{Sigmaeph}, we find that to linear order in $\omega\ll\omega_0$, $\Re e \Sigma(\omega) = - \lambda \,\omega$. Therefore, $\lambda = \frac{v_f}{V_L}-1$ (see also \cite{Chubukov2004}). According to the normal state data ($T=115 \ \text {K}$) from \cite{Kaminski2001,Vig2015,Norman2007} (since $T\ll\omega_0$, this zero temperature formula still applies), $V_L=1.47$eV\,$\AA$ and $v_f=2.7$eV\,$\AA$, yielding $\lambda=0.84$. In principle, one might argue for the larger value of $v_f\sim 5.4$ eV\,$\AA$ from the ARPES observed width of the band \cite{Gweon2011}, leading to 
$\lambda \sim 2.67$, a very  high value indeed. However, we will assume, with several authors of the Boson-coupling models, that  the smaller estimate is overall  more reasonable.
Using these experimentally relevant values, in \figdisp{EDC_MDC_FG_T115}, we plot $\omega - \Re e \Sigma(\omega)$ and $-\Im m \Sigma(\omega)$ vs. $\omega$ (left panel), as well as the MDC and EDC dispersions (middle panel), and the EDCs at several representative momenta (right panel). Due to the higher value of $T$, the self-energy curves have been rounded out somewhat as compared to \figdisp{EDC_MDC_FG}, but retain the same features. We see that the EDC dispersion once again follows the MDC dispersion (closest to zero frequency) in the first two momentum regions, until it (nearly) flattens out in the third region, where the peak is pinned to the phonon frequency, $-\omega_0$, in the corresponding EDCs. As the momentum is increased such that the height of this peak shrinks below the height of the hump, the EDC dispersion jumps discontinuously down from the phonon frequency, to the MDC dispersion. Consequently, we see that the velocities of the MDC and EDC dispersion coincide above the kink; i.e.  $V_H=V_H^*$.

\FloatBarrier

\begin{figure}[htp]
\begin{center}
\includegraphics[width=0.33\columnwidth]{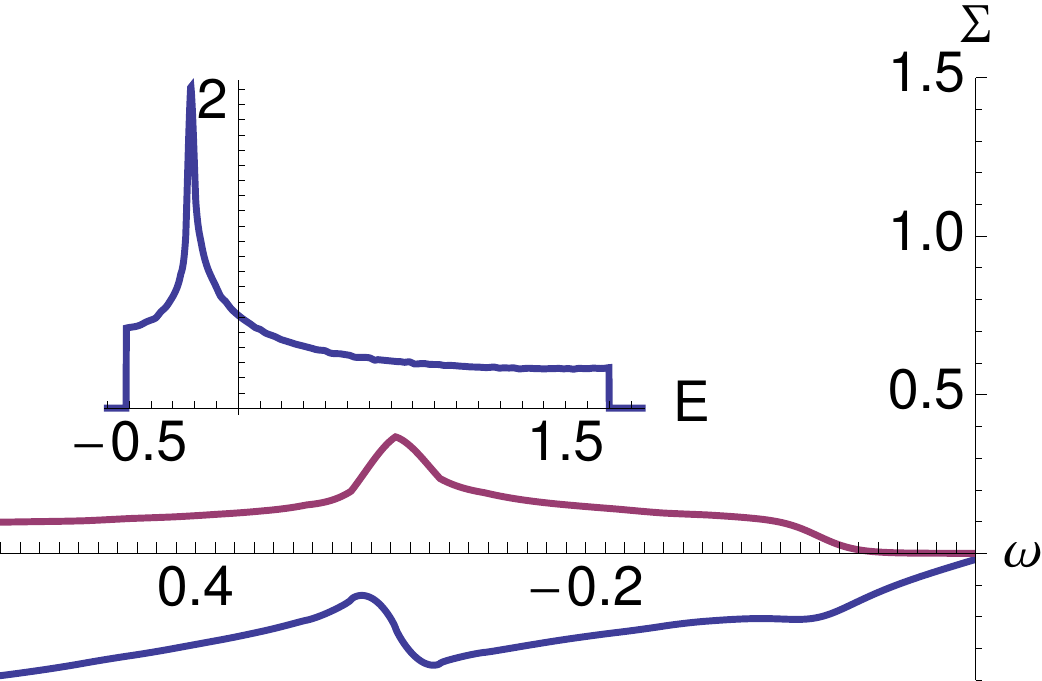}
\includegraphics[width=0.32\columnwidth]{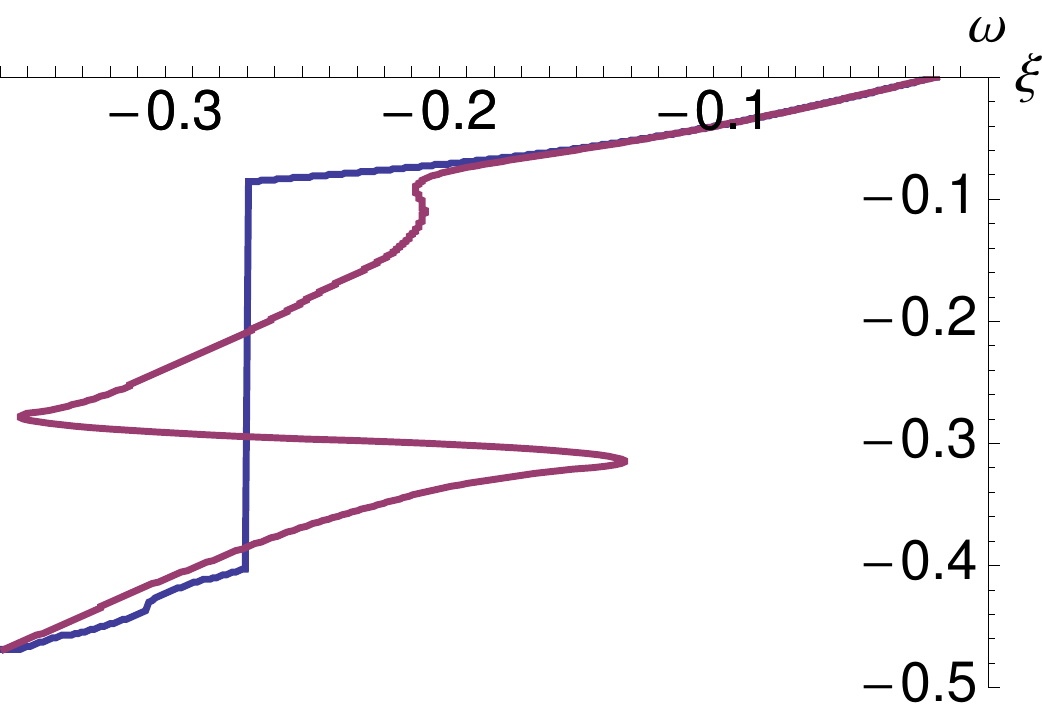}
\includegraphics[width=0.33\columnwidth]{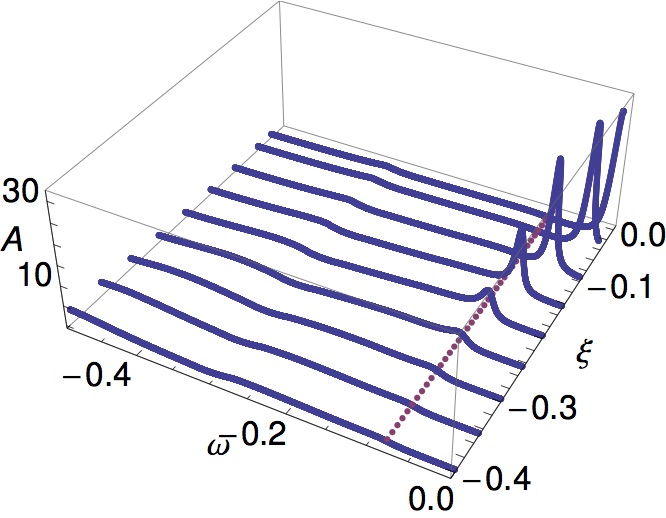}
\caption{We explore the effects of using the full frequency-dependence of the density of states in \disp{Sigmaeph}, with $\lambda=0.84$ and $T= 115 \ \text{K}$. Due to the functional form of the density of states (displayed as an inset in the left panel), the MDC dispersion acquires two additional branches which yield large frequency values. Below the low-energy kink momentum, the EDC dispersion follows the lowest-frequency branch of the MDC dispersion. Above the low-energy kink momentum, the EDC dispersion initially stays pinned to the phonon frequency, until it discontinuously jumps onto the highest-frequency branch of the MDC dispersion ($V_H=V_H^*$). A noticeable hump also develops at high-frequencies, in the corresponding EDCs.} 
\label{EDC_MDC_FG_DOS} 
\end{center}
\end{figure}

We now examine how these results are affected by retaining the full frequency-dependence of the density of states in \disp{Sigmaeph}. Just as was done in \cite{Vig2015}, we use the dispersion tb2 from \cite{Norman2007}. In this case, $\varepsilon_f=0$ and $N(\varepsilon_f) = 0.61 \ \text{eV}^{-1}$. Retaining the same values of $T=115 \ \text {K}$ and $\lambda=0.84$, we set $g=0.23 \ \text{eV}$ in \disp{Sigmaeph}. We also set $\mu\approx\varepsilon_f=0$. In \figdisp{EDC_MDC_FG_DOS}, we plot $\omega - \Re e \Sigma(\omega)$ and $-\Im m \Sigma(\omega)$ vs. $\omega$ (left panel), as well as the MDC and EDC dispersions (middle panel), and the EDCs at several representative momenta (right panel). Due to the functional form of the density of states (see the inset of the left panel), the MDC dispersion acquires two additional branches which yield large frequency values. In the first two momentum regions (below the low-energy kink momentum), the EDC dispersion follows the lowest-frequency branch of the MDC dispersion. As the momentum increases into the third region (above the low-energy kink momentum), the peak stays pinned to the phonon frequency in the corresponding EDCs. Moreover, since $|\Im m \Sigma\left(E(\xi)\right)| \gg |\Im m \Sigma(-\omega_0)|$, where $E(\xi)$ can be any branch of the MDC dispersion, the EDC dispersion stays pinned to the phonon frequency as well. As the momentum is increased further and the height of the peak decreases sufficiently, the EDC dispersion jumps discontinuously onto the highest-frequency branch of the MDC dispersion, since this is the one with the smallest value of $|\Im m \Sigma\left(E(\xi)\right)|$, and hence $V_H=V_H^*$. This small value of $|\Im m \Sigma\left(E(\xi)\right)|$ leads to a noticeable hump at high-frequencies in the corresponding EDCs.

\FloatBarrier

Thus far, we have considered only free electrons coupled to a Boson mode. We now include electron-electron correlations. Following \cite{Shastry_2011_Anatomy}, we assume that 
\beq
\Im m \Sigma_{el-el}(\omega) = - \frac{(\tau ^2+\omega^2)}{\Omega _0}\exp\left[\frac{-(\tau ^2+\omega ^2)}{\nu _0^2}\right]  -  \ \eta, \nn\\
\label{Sigma_el_el}
\eeq
where $\Sigma_{el-el}(\omega)$ is the self-energy due only to electron-electron correlations, $\tau \equiv \pi k_B T$, $T= 115 \ \text{K}$, $\Omega_0=.14 \ \text{eV}$, $\nu_0=.5 \  \text{eV}$, and we set $\eta = .01 \ \text{eV}$. This phenomenological form for $\Im m \Sigma_{el-el}(\omega)$ reproduces the correct Fermi-liquid behavior at low frequencies, and extrapolates to high frequencies in a reasonable way. Furthermore, we assume a flat band for $\varepsilon_k$ of bandwidth $W$, i.e $N(E) = \frac{1}{W} \Theta(\frac{W}{2}-|E|)$, and set $\mu\approx\varepsilon_f=0$. Retaining the same values of $N(\varepsilon_f) = 0.61 \ \text{eV}^{-1}$ and $\lambda=0.84$ as before, yields the values $W=1.64 \ \text{eV}$ and $g=0.23 \ \text{eV}$. The self-energy is now given by the sum $\Sigma(\omega) = \Sigma_{el-el}(\omega) + \Sigma_{el-ph}(\omega)$, where the imaginary part of the latter term is 
\beq
\Im m \Sigma_{\text{el-ph}}(\omega) &=& - \pi g^2 \sum_{\pm} A_{\text{el-el,loc}}(\omega\pm\omega_0)\times\nn\\
&&\left[f^{\mp}(\omega\pm\omega_0)+n(\omega_0)\right],
\label{Sigma_eph_FL}
\eeq
while the real part is as usual given by applying the Hilbert transform to \disp{Sigma_eph_FL}. Here, $A_{\text{el-el,loc}}(\omega) = \frac{1}{N_s}\sum_k A_{\text{el-el}}(\vk,\omega)$, where $A_{\text{el-el}}(\vk,\omega)$ is given by \disp{A_ph} with the substitution $\Sigma(\omega) \to \Sigma_{\text{el-el}}(\omega)$. \disp{A_ph} continues to express $A(\vk,\omega)$ in terms of $\Sigma(\vk,\omega)$, where both objects now include electron-electron and electron-phonon correlations.

\begin{figure}[htp]
\begin{center}
\includegraphics[width=0.33\columnwidth]{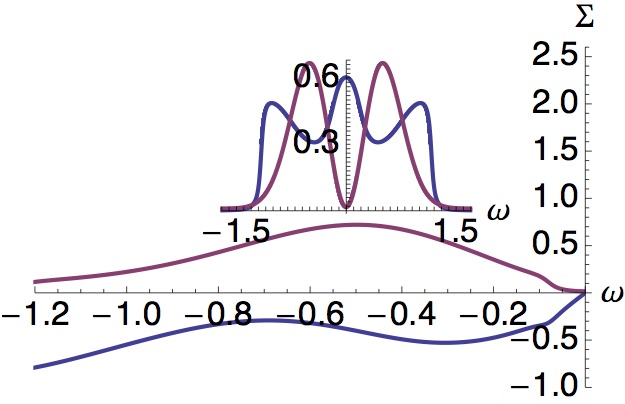}
\includegraphics[width=0.32\columnwidth]{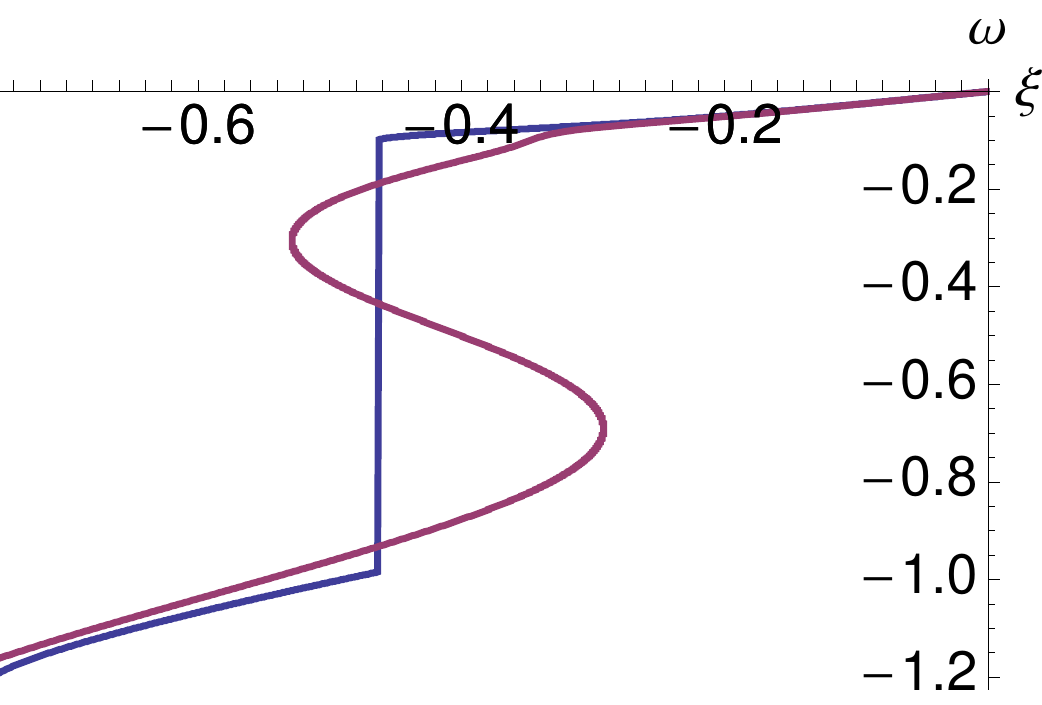}
\includegraphics[width=0.33\columnwidth]{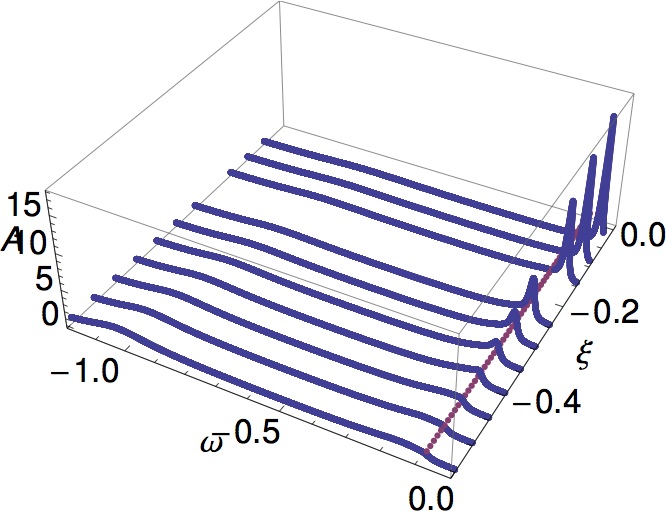}
\caption{We explore the effects of Fermi-liquid-like electron-electron correlations (\disp{Sigma_el_el}), with $\lambda=0.84$ and $T=115 \ \text{K}$. Due to the functional form of the self-energy, $\Sigma_{el-el}(\omega)$ (both $-\Im m \Sigma_{el-el}(\omega)$ and $A_{\text{el-el,loc}}(\omega)$ are displayed as an inset in the left panel), the highest-frequency branch of the MDC dispersion yields very large values of the frequency. Below the low-energy kink momentum, the EDC dispersion follows the lowest-frequency branch of the MDC dispersion. Above the low-energy kink momentum, the EDC dispersion initially stays pinned to the phonon frequency, until it discontinuously jumps onto the highest-frequency branch of the MDC dispersion ($V_H=V_H^*$). This is also reflected in the corresponding EDCs, which acquire a hump at high-frequencies.} 
\label{EDC_MDC_FL} 
\end{center}
\end{figure}

In \figdisp{EDC_MDC_FL}, we plot $\omega - \Re e \Sigma(\omega)$ and $-\Im m \Sigma(\omega)$ vs. $\omega$ (left panel), as well as the MDC and EDC dispersions (middle panel), and the EDCs at several representative momenta (right panel), from this calculation. Due to the specific form of the self-energy, $\Sigma_{el-el}(\omega)$ (both $-\Im m \Sigma_{el-el}(\omega)$ and $A_{\text{el-el,loc}}(\omega)$ are displayed as an inset in the left panel), the highest-frequency branch of the MDC dispersion yields very large values of the frequency. Just as in the cases considered above, for momentum $|\xi|$ below the low-energy kink momentum, the EDC dispersion follows the lowest-frequency branch of the MDC dispersion, $E_l(\xi)$. As the momentum $|\xi|$  is increased above the low-energy kink momentum, the rapid increase in $|\Im m \Sigma\left(E_l(\xi)\right)|$ causes the peak in the EDC as well as the EDC dispersion to stay pinned to the phonon frequency. As the momentum is increased further, $|\Im m \Sigma\left(E_h(\xi)\right)|$ becomes comparable to $|\Im m \Sigma(-\omega_0)|$, where $E_h(\xi)$ is the highest-frequency branch of the MDC dispersion. At this point, the EDC dispersion jumps discontinuously from the phonon frequency onto the highest-frequency branch of the MDC dispersion, and hence $V_H=V_H^*$. This is also reflected in the corresponding EDCs, which acquire a hump at high-frequencies.

\FloatBarrier

In conclusion, we find that in all of the above cases of electrons interacting with a Boson mode, the EDCs  are characterized by three signatures:  (1)  a peak pinned to the Boson-frequency over a large range of momentum, (2) the EDC dispersion jumps discontinuously from the Boson-frequency onto (the highest-frequency branch of) the MDC dispersion, and (3) $V_H=V_H^*$. These three features are  jointly present for most parameters explored, and may be viewed as the signatures of   kinks  produced by  this mechanism.

\section{(III)  Extremely Correlated Fermi liquid  theory  of kinks}
In this section we present the theoretical  details of the  ECFL calculation of kinks.
 We first show the results of a low energy and momentum expansion of the ECFL Greens function in terms of a few parameters.    Earlier studies \cite{ECFL-DMFT,ECFL-AIM,Hansen-Shastry} show  that the two self energies $\Phi,\Psi$ of the ECFL theory are to a large extent similar to the self energies of a standard intermediate coupling Fermi liquid, and yet  due to their specific combination that occurs in \disp{twin-self} and \disp{twin-self-2} end up providing a  non trivial resulting theory. Indeed in \refdisp{ECFL-DMFT} a similar low energy expansion  in high dimensions,  was tested successfully against the numerical results of the Dynamical Mean Field Theory (DMFT). It should be noted that the DMFT theory is designed for high dimensions, where the momentum dependence of the Dyson self energy and $\Psi$ self energy of the ECFL theory  drops out. In this section we allow for momentum dependence of both self energies in the ECFL formalism, this is in-fact the only distinction between the present expansion and that in \refdisp{ECFL-DMFT}. We see below that this momentum dependence is essential for describing the low energy kinks in the occupied part of the  ARPES spectrum.

\subsection{  Low energy expansion of the ECFL theory}
We start with the ECFL  Greens function $\G$  expressed in terms of the auxiliary Greens function $\GH$ and the caparison function $\widetilde{\mu}$  \refdisp{Shastry-2011} and  \refdisp{Shastry2014}, we write
\beq
\G(\vec{k}, i\omega)= \GH(\vec{k}, i\omega) \times \widetilde{\mu}(\vec{k}, i\omega),  \label{twin-self}
\eeq
and with  the latter expressed in terms of the two self energies  $\Phi(\vec{k}, i \omega_n),\Psi(\vec{k}, i \omega_n)$ as:
\beq
\widetilde{\mu}(\vec{k}, i \omega_n)&=& 1- \frac{n}{2} + \Psi(\vec{k}, i \omega_n) \label{caparison} \\
\GH^{-1}(\vec{k}, i \omega_n)&=& i \omega_n + \chem - (1-\frac{n}{2} ) \varepsilon_k   - \Phi(\vec{k}, i \omega_n), \nn \\ \label{twin-self-2}
\eeq
where $n$ is the electron number per site, $\omega_n = (2 n+1) \pi/\beta$ the Matsubara frequency, which  we  analytically continue $i\omega \to \omega+ i 0^+$. Let us define $\hat{k}$ as the {\em normal deviation} from the Fermi surface  i.e. $\hat{k}= (\vec{k}- \vec{k}_F). \vec{\nabla}\varepsilon_{k_F}/|\vec{\nabla}\varepsilon_{k_F}|$. Our first objective is to Taylor expand these equations for small $\omega$ and $\hat{k}$, as explained above. We carry out a low frequency expansion  as follows:
\beq
1- \frac{n}{2} +\Psi(\vec{k},\omega) &=& \alpha_0 + c_\Psi (\omega + {\vv_\Psi} {\kkk} ) 
 + i {\cal R}/\gamma_\Psi + {\cal O}(\omega^3),
\nn \\   \label{psiex}
\eeq
where  the frequently occurring Fermi liquid function
${\cal R}= \pi \{ \omega^2+ (\pi k_B T)^2\}$,  $v_f =(\partial_k \varepsilon_k)_{k_F}$ is the {\em bare} Fermi velocity, and the four  parameters  $\alpha_0, c_\Psi, {\vv_\Psi}, \gamma_\Psi  $ are coefficients in the Taylor expansion having suitable dimensions. Similarly we expand the auxiliary Greens function
\beq
\GH^{-1}(k,\omega)   =   (1+ c_\Phi ) \left(  \omega - \vv_\Phi \kkk + i {\cal R} /\Omega_\Phi +{\cal O} (\omega^3) \right), \nn \\ 
\label{phiex}
\eeq
where we have added another three coefficients in the Taylor expansion $ c_\Phi, \vv_\Phi, \Omega_\Phi $.  

To  carry out this reduction we  first trade the two parameters $c_\Psi, \gamma_\Psi$ in favor of  parameters $\Omega_\Psi$ and $s$ by defining $
c_\Psi= \frac{\alpha_0}{\Omega_\Psi}$  and
$\gamma_\Psi = \frac{s \Omega_\Phi}{c_\Psi}$, 
 where  the dimensionless  parameter $  0 \leq s \leq 1$.
With these expansions and the  quasiparticle weight determined in terms of the expansion parameters as
$
Z= \frac{\alpha_0}{1+ c_\Phi}$, we find 
\beq
\G=\frac{Z}{\Omega_\Psi} \frac{\Omega_\Psi + \omega + \vv_\Psi \kkk + i {\cal R} /( s \Omega_\Phi)}{\omega- \vv_\Phi \kkk + i {\cal R}/\Omega_\Phi} \label{Gagain}.
\eeq

Using $A(\hat{k},\omega) = - \frac{1}{\pi} \Im m \, \G$  we find the spectral function
\beq
A(\hat{k},\omega) &=& \frac{Z}{\pi} \frac{ \frac{{\cal R}}{\Omega_\Phi}}{(\omega - \vv_\Phi \kkk)^2+(\frac{{\cal R}}{\Omega_\Phi})^2} \times \widetilde{\mu}_c(\hat{k},\omega)  \nn \\ 
\label{spectral-function-again}
\eeq
Here the  caparison {\em factor}, (not to be confused with  the caparison {\em function} in \disp{twin-self}),  is found as 
 \barray
 \widetilde{\mu}_c(\hat{k},\omega) & =& 1- \xi(\hat{k},\omega) \nn \\
 \xi(\hat{k},\omega)& =& \frac{1}{\Delta_0} (\omega- \vv_0 \kkk ) \label{caparison}
 \earray
 In  \disp{caparison} we have introduced   two composite parameters
\beq
\Delta_0  =  \frac{s}{1-s} \Omega_\Psi \label{delta-again}, \;\;\mbox{and} \;\;\; \vv_0 =  \frac{1}{1-s} \vv_\Phi + \frac{s}{1-s} \vv_\Psi \label{velocity-again} . 
\eeq
This procedure eliminates the {\em three}  old parameters $s$,  $\Omega_\Psi$ and $\vv_\Psi$ in favor of the {\em two}  emergent energy scale $\Delta_0$ and velocity $\vv_0$.

It is interesting to count the reduction in the number of free parameters from the starting value of seven in \disp{psiex} and \disp{phiex}. Already in \disp{Gagain} we have a reduction to six, since the quasiparticle weight $Z$  combines two of the original parameters. Since \disp{velocity-again} subsumes three parameters into two,   the spectral function in \disp{spectral-function-again} contains only five parameters: the two velocities $\vv_0 \, v_f, \vv_\Phi \, v_f,$ and the two energies $  \Omega_\Phi, \Delta_0$,  in addition to the overall scale factor  $Z$.

We will see below that the parameters that are  measurable  from energy dispersions 
are best expressed in terms of certain  combinations of the velocities. In order to make the connection with the experiments close, we will redefine the two velocities in terms of an important dispersion velocity at the lowest energies $V_L$ and a dimensionless ratio $r$, on using the definitions:
\beq
\vv_\Phi \, v_f  &=& V_L  \nn \\
\vv_0 \, v_f &=&  r \times V_L . \label{rediff-1}
\eeq

In order to account for the difference between laser ARPES and  synchrotron AREPS having different incident photon energies, we will make two phenomenological  modifications in \disp{spectral-function-again} following \refdisp{Gweon2011}
\beq
{\cal R}(\omega)/\Omega_\Phi \to {\cal R}(0)/\Omega_\Phi=  \pi \{ \pi k_B T \}^2/\Omega_\Phi + \eta  \equiv   \Gamma_0
\eeq
where  $\eta$ represents an elastic energy from impurity scattering,  dependent upon the energy of the incident photon in the ARPES experiments. In the spirit of a low energy expansion   ${\cal R}$ is evaluated at $\omega=0$. Thus  $\Gamma_0$ is a T dependent constant, which subsumes the two parameters $\eta$ and $\Omega_\Phi$, and thus the total parameter count is still five. 
Secondly for extension to higher energies,  we  ``renormalize'' the parameter $\xi$ in \disp{caparison} according to a recently discussed prescription following from a theoretical calculation \refdisp{Shastry-2016} as
$\widetilde{\mu_c}\to \{ 1-  \frac{\xi}{\sqrt{1+ c_a \xi^2} }  \},$
where $c_a\sim 5.4$ near  optimum doping $\delta\sim 0.15$ as estimated recently. This correction ensures that the caparison factor exhibits the correct linear behavior for small $\xi$, and remains positive definite at high energies.  Thus we write the spectral function in terms of the new variables as
\beq
A(\vec{k},\omega)= \frac{Z}{\pi} \frac{\Gamma_0}{(\omega- V_L \, \hat{k})^2 + \Gamma_0^2} \times  \{ 1-  \frac{\xi}{\sqrt{1+ c_a \xi^2} }  \}, \nn \\ \label{spectral-function}
\eeq
with  $ \xi= \frac{1}{\Delta_0} (\omega- r \, V_L \, \hat{k} )$.
 We should keep in mind  that  these expressions follow from   a  low energy expansion, and  is limited to small $\hat{k}$ and $\omega$,  so that   the dimensionless variable $|\xi|_{max} \sim {\cal O}(1)$. Microscopic calculations of all these parameters is possible in the ECFL theory.  One important parameter is the energy scale $\Delta_0$ which is found to be much reduced from the band width, due to extremely strong correlations. A related energy  is the effective Fermi liquid temperature scale where the $T^2$ dependence of the resistivity gives way to a linear dependence. This scale is estimated in the limit of large dimensions from \refdisp{Shastry-2016} to be as low as 45 K near optimum doping, i.e. much reduced from naive expectations.

For the present purposes we take a different track, we  note that the  ARPES fits are overdetermined, so that we can 
determine the few parameters of the low energy theory from a fairly small subset of measurements.    The five final (composite)   parameters defining the spectral function \disp{spectral-function}  are $Z, V_L, r ,  \Delta_0, \Gamma_0$, where $c_a\sim 5.4$. Of these $Z$ is multiplicative, it is only needed for getting the absolute scale of the spectral function, and $c_a$ does not play a significant role near  zero energy, it is required only at high energies. Thus the spectra relevant to  EDC and MDC will require only {\em four} parameters $V_L, r ,  \Delta_0, \Gamma_0$. These suffice to determine the low energy theory and thus to  make a large number of predictions; i.e.   implying non trivial relationships amongst observables. Many of the predictions rely only on the overall structure of the theory and not its details.

\subsection{The EDC and MDC  dispersion relations and kinks}

Starting from \disp{spectral-function}, we can compute the energy dispersions for MDC (varying $\hat{k}$ while keeping $\omega$ fixed) and the EDC spectra (varying  $\omega$ while keeping $\hat{k}$ fixed). In terms of a  momentum type variable
\beq Q(\hat{k})= \Delta_0 +  (r-1) \hat{k}\, V_L \,  \label{Qdef} \eeq 
 we can locate the peaks of \disp{spectral-function} using elementary calculus since $c_a$ only plays a role at high energies, we set $c_a\to 0$ when performing the extremization and find the MDC dispersion
 \beq
 E(k)&=& \frac{1}{2-r} \left(  \hat{k} \, V_L  + \Delta_0 - \sqrt{r (2-r) \, \Gamma_0^2  + Q^2} \right), \; \nn \\  \label{MDC} 
 \eeq
 and the EDC dispersion
 \beq
 E^*(k)&=&  \left( r \; \hat{k} \,  V_L + \Delta_0 - \sqrt{\Gamma_0^2+ Q^2} \right).\label{EDC}
 \eeq
Using these two dispersions and expanding them in different regimes, we can extract all the parameters of the kinks.

\subsubsection{ Kink momentum }
As explained in the main paper, when we set $T=0=\eta$ so that $\Gamma_0=0$, both the EDC and MDC dispersions contain an ideal kink at the kink momentum. Therefore, using Eqs. (\ref{MDC}) and (\ref{EDC}), the condition  $Q=0$ locates the kink momentum for both dispersions:
 \beq
 \hat{k}_{kink} = \frac{\Delta_0}{ (1-r) V_L}, \label{kinkmomentum}
 \eeq
 it corresponds to occupied momenta, i.e.  $\hat{k}_{kink} v_f<0$, provided that $r >1$.
We thus can express
$
\Delta_0 = \hat{k}_{kink} \; V_L (1-r)$, enabling us to 
 usefully rewrite
\beq
Q= (r-1) \,V_L \, (\hat{k}- \hat{k}_{kink}) = \Delta_0 \, \{ 1- \frac{\hat{k}}{\hat{k}_{kink}}  \} \label{Q-Delta}. 
\eeq
As required by the ideal kink, $Q$ changes sign at the kink momentum,
\beq
\mbox{sign}(Q) = \mbox{sign}(\hat{k}- \hat{k}_{kink}).
\eeq

\subsubsection{Ideal Kink energies: T=0 }
Using \disp{MDC} and \disp{EDC}, in conjunction with \disp{kinkmomentum}, the ideal kink energy is the same for both dispersions, and is given by
\beq
E^{ideal}_{kink}= - \frac{1}{r-1} \Delta_0 \label{ideal}.
\eeq
We can also usefully estimate this ideal kink energy  from the asymptotic velocities in the far zone,  as explained in the main paper.

\subsubsection{ The non-ideal i.e. $T>0$  kink energy}

The EDC and MDC kink energies  for the non-ideal case can be viewed in a couple of ways. We have argued in the main paper that these are best defined by fixing the momentum $\hat{k}= \hat{k}_{kink}$ and reading off the energy at this value. This is an unambiguous method independent of the detailed shape of the kink,  since it only requires knowledge of $\hat{k}_{kink}$, which can be found from an asymptotic measurement as we have argued in the main paper. We can  put $Q=0$ and $\hat{k} \to \hat{k}_{kink}$ in \disp{EDC} and \disp{MDC} and read off the kink energies:
 \beq 
 E^{EDC}_{kink}& =&  E^{ideal}_{kink} - \Gamma_0, \label{EDC-kink}\\
 E^{MDC}_{kink} &=& E^{ideal}_{kink}- \Gamma_0 \sqrt{\frac{r}{2-r}}. \label{MDC-kink}
 \eeq
We observe that the MDC  kink  energy is real provided $2 \geq r \geq 1$. Note also that at $T=0$ and $\eta=0$, the two energies both reduce to the ideal kink energy. 

\subsubsection{ The ideal   energy dispersions }
At $T=0$ or for $|Q| \gg\Gamma_0$,  the two dispersions \disp{EDC} and \disp{MDC}   become: 
  \beq
 E^*(k) \sim \left[ r - (r-1) \, \mbox{sign}(\hat{k}-\hat{k}_{kink} )\right]  \hat{k} \, V_L \ + \ 2 \Delta_0 \Theta(\hat{k}_{kink}-\hat{k})
 \eeq
 and
\beq
 E(k) \sim \frac{1}{2-r} \left[ 1 - (r- 1) \, \mbox{sign} (\hat{k}-\hat{k}_{kink} ) \right]  \hat{k} \, V_L \ + \ \frac{2 \Delta_0}{2-r} \Theta(\hat{k}_{kink}-\hat{k})
. \nn \\ 
 \eeq 
 The  velocities in the asymptotic regime $|\hat{k}|\gg \hat{k}_{kink}$ can be found from the slopes of these, and are therefore temperature-independent.  For $\hat{k} \gg \hat{k}_{kink}$  we get the ``low'' velocities
\beq
 \frac{ dE(k)}{d\hat k} &=&  V_L \nn \\
\frac{ dE^*(k)}{d\hat k} &=&  V^*_L=V_L 
  \label{V-low-0}
\eeq
and thus the EDC and MDC velocities are identical.
For $\hat{k}\ll \hat{k}_{kink}$ we get the ``high'' velocities
\beq
V_H&=& \frac{ d E(k)}{d\hat{k}} = \frac{r}{2-r} V_L, \label{V-high-MDC} \\
V_H^*&=&  \frac{ d E^*(k)}{d\hat{k}}= (2 r-1) V_L. \label{V-high-0}
\eeq
We may cast \disp{V-high-0} into an interesting  form
\beq
V_H^*= \left\{ \frac{ 3 V_H - V_L}{V_H + V_L} \right\} V_L,
\eeq
it is significant since  the EDC spectrum velocity  is exactly determined in terms of the two MDC spectrum velocities. It is also a testable result, we show elsewhere in the paper how this compares with known data. Note that the four independent parameters $V_L, r ,  \Delta_0, \Gamma_0$ alluded to in the discussion below \disp{spectral-function}, can be determined from the directly measurable parameters $V_L,V_H,\hat{k}_{kink},\Gamma_0$ (\ref{V-high-MDC},\ref{kinkmomentum},\ref{gamma0}). Therefore, either set of parameters gives complete knowledge of the EDC and MDC dispersions, as well as the spectral function (up to an overall scale).

\subsubsection{ Near Zone: Corrections to Energy dispersion due to finite T.}
In the regime dominated by finite T and  effects of $\eta$ the elastic scattering parameter, we can also perform an    expansion in the limit when $|Q|\ll \Gamma_0$, using \disp{MDC} and \disp{EDC}. The  the first few terms are
\beq
E(k)&=& \frac{\Delta_0}{1-r} - \sqrt{\frac{r}{2-r}}\, \Gamma_0 + \frac{V_L}{2-r} (\hat{k}- \hat{k}_{kink}) \nn \\
&& - \frac{(1-r)^2}{2 \sqrt{r (2-r)^3}} \, \frac{V^2_L}{\Gamma_0} (\hat{k}- \hat{k}_{kink})^2 + \ldots \label{MDC-quadratic}
\eeq
Similarly for the EDC dispersion
\beq
E^*(k)&=& \frac{\Delta_0}{1-r} -  \Gamma_0 +  r {V_L} (\hat{k}- \hat{k}_{kink}) \nn \\
&& - \frac{(1-r)^2}{2 } \, \frac{V^2_L}{\Gamma_0} (\hat{k}- \hat{k}_{kink})^2 + \ldots
\eeq
These formulas display  a shift in the energies due to $\Gamma_0$ and also a  $\Gamma_0$ dependent curvature. 
Since the regime of this expansion, $|Q|< \Gamma_0$ is different from  that of the expansion
in \disp{V-high-0} and \disp{V-low-0}, we note that velocities are different as well. Thus one must be careful about specifying the regime for using the velocity formulae. 

Let us note that in this regime $|Q|<\Gamma_0$ the two  dispersions differ, with the EDC higher.
\beq
&&E^*(k)-E(k)= \{\sqrt{\frac{r}{2-r}} -1 \}  \Gamma_0 \nn \\
&&- \frac{(1-r)^2}{2-r} {V_L} (\hat{k}- \hat{k}_{kink})+\ldots \label{difference}
\eeq
This equation gives a prescription for estimating $\Gamma_0$ in cases where the other parameters are known. Alternatively  in the MDC dispersion we expect to see a curvature only near the location of the kink, this is sufficient to fix $\Gamma_0$: 
from \disp{MDC-quadratic}
\beq
\frac{d^2 E(k)}{d \hat{k}^2}&=&- \frac{(r-1)^2}{ \sqrt{r(2-r)^3}} \; \frac{V^2_L }{\Gamma_0}.
\eeq
The curvature   $ \frac{d^2 E(k)}{d \hat{k}^2}$ can be estimated from the experimental data  to provide an estimate of  $\Gamma_0$.

\subsection{ The Dyson self energy } 
 For completeness we present the low energy expansion of the Dyson self energy, which  gives rise to the spectral function in \disp{spectral-function}.  We may   define the Dyson self energy from 
\beq
 \Sigma_D&=&\omega+ \chem- \varepsilon_k - \G^{-1} \label{dyson-self}
 \eeq
 Using \disp{Gagain} we obtain
 \beq
 \Im m \, \Sigma_D= - \frac{1}{Z} \frac{{\cal R}}{\Omega_\Phi} \frac{1 - \frac{1}{\Delta_0}(\omega - \vv_0 \kkk ) }{\{1+(\omega+ \vv_\Psi \kkk)/\Omega_\Psi \}^2+ \frac{{\cal R}^2}{s^2\Omega^2_\Phi \Omega^2_\Psi} }  \nn \\ \label{imsigmadyson}
 \eeq
 The corresponding real part is given by 
 \beq
  \Re e \, {\Sigma}_D&=& \chem-\chem_0 + \omega - \kkk  \nn \\
  &&- \frac{1}{Z} \frac{(\omega - \vv_\Phi \kkk) + \frac{1}{\Omega_\Psi} q_2}{{\{1+(\omega+ \vv_\Psi \kkk)/\Omega_\Psi\}^2+ \frac{{\cal R}^2}{s^2\Omega^2_\Phi \Omega^2_\Psi} } } \nn\\
  q_2&=&(\omega+ \vv_\Psi \kkk)(\omega - \vv_\Phi \kkk)+ \frac{{\cal R}^2}{s \Omega_\Phi^2}. \label{resigmadyson}
  \eeq
The $q_2$ term is  quadratic (or  higher) in the small variables $\omega,\kkk$, however  these small terms are needed if we want to reproduce exactly \disp{spectral-function-again}.

\subsubsection{Useful identities and some Fermi Liquid parameters.}
We  list  a few useful identities relating the  various parameters
\beq
\Omega_\Psi&=& \frac{1-s}{s} \, \Delta_0, \nn \\
s &=& \frac{\Delta_0}{\Delta_0+ \Omega_\Psi} \nn \\
\vv_0&=& \frac{\vv_\Phi+ s \, \vv_\Psi}{1-s} = r \,\vv_\Phi \nn \\
\vv_\Psi&=& \frac{r-1-r s }{s} \, \vv_\Phi \nn \\
r-1&=& \frac{\Delta_0}{\Omega_\Psi} \left( 1+ \frac{\vv_\Psi}{\vv_\Phi} \right) \label{identities}
\eeq

Let us note the Fermi liquid renormalizations from \disp{dyson-self}
\beq
 \frac{d \Sigma_D}{d \hat{k}}   \bigg|_{FS} &=&  (  \frac{V_L}{Z} -v_f ) \nn \\
 \frac{d \Sigma_D}{d \omega}  \bigg|_{FS} &=&  (1 - \frac{1}{Z} ) \nn \\
\eeq
Therefore we write the Fermi liquid  mass enhancement that determines the heat capacity as:
\beq
\frac{m}{m^*} = Z \left\{1+ \frac{1}{v_f}  \frac{d \Sigma_D}{d \hat{k}}  \bigg|_{FS}  \right\} =V_L/v_f = \vv_\Phi.
\eeq
Thus $\vv_\Phi$ is the inverse mass enhancement factor, obtainable  from the ratio of the heat capacity and the bare density of states. In this model  we note that $\vv_\Phi$ is not obliged to vanish as $Z$ near the half filled limit $n\to1$, but may be a finite number of $O(1)$. This is  unlike the Brinkman Rice ``heavy metal' type behavior $m/m^* \propto Z$, which is prototypical of theories with a momentum independent self energy.

Finally we note that  the condition for the kink to occur is, we recall, $r>1$. From \disp{identities} we see that this requires a finite $\Omega_\Psi$ (so that $1>s>0$). We also need $\Delta_0 >0$ and  $\left( 1+ \frac{\vv_\Psi}{\vv_\Phi} \right)>0$.